\PassOptionsToPackage{svgnames,table,xcdraw,rgb}{xcolor}

\documentclass[10pt,journal,compsoc]{IEEEtran}
\usepackage{hyperref}
\usepackage{amsmath,amsfonts}
\usepackage{url}
\usepackage{verbatim}
\usepackage{graphicx}
\usepackage{cleveref}
\usepackage{cite}

\usepackage{booktabs}
\usepackage{multirow}
\usepackage{todonotes}

\usepackage[switch]{lineno}

\renewcommand\appendixname{Supplemental Material}
\renewcommand\appendix{\section*{\appendixname}\setcounter{section}{0}}

\graphicspath{{figs/}{figures/}{pictures/}{images/}{./}} %

\usepackage{tabu}                      %
\usepackage{booktabs}                  %
\usepackage{lipsum}                    %
\usepackage{mwe}                       %
\usepackage{csquotes}
\usepackage{mathptmx}                  %
\usepackage{amsmath}

\usepackage{rotating}
\usepackage{tabularx}
\usepackage{enumitem}

\usepackage[framemethod=TikZ]{mdframed}

\usepackage{url}
\AtEndPreamble{\usepackage{tikz}}
\usepackage{listings}
\definecolor{color1}{HTML}{5778a4}
\definecolor{color2}{HTML}{e49444}
\definecolor{color3}{HTML}{6FBD6F}
\definecolor{color4n}{HTML}{E64A7B}
\definecolor{color4}{HTML}{E64A7B}

\newcommand\chatgpt[1][]{\tikz[overlay, remember picture]{
    \node[inner sep=0pt, anchor=text](chatgptnode){\phantom{GPT-4o}};
    \draw[color1, line width=2pt] 
      ([yshift=-0.5pt]chatgptnode.south west) -- 
      ([yshift=-1pt,xshift=2pt]chatgptnode.south east);
  }%
  {\mbox{GPT\mbox{-}4o}}\xspace%
}

\newcommand\gemini[1][]{\tikz[overlay, remember picture]{
    \node[inner sep=0pt, anchor=text] (gemininode) {\phantom{Gemini-2.5}};
    \draw[color2, line width=2pt] 
      ([yshift=-1pt]gemininode.south west) --
      ([yshift=-1pt]gemininode.south east);
  }\mbox{Gemini\mbox{-}2.5}\xspace%
}

\newcommand\qwen[1][]{%
  \tikz[overlay, remember picture]{
    \node[inner sep=0pt, anchor=text] (qwennode) {\phantom{Qwen2.5}};
    \draw[color4, line width=2pt] 
      ([yshift=-1pt]qwennode.south west) --
      ([yshift=-1pt]qwennode.south east);
  }\mbox{Qwen2.5}\xspace%
}

\newcommand\human[1][]{%
  \tikz[overlay, remember picture]{
    \node[inner sep=0pt, anchor=text] (humannode) {\phantom{human~subjects}};
    \draw[color3,  line width=2pt] 
      ([yshift=-1pt]humannode.south west) --
      ([yshift=-1pt,xshift=1pt]humannode.south east);
  }\mbox{human~subjects}\xspace%
}

\newcommand\humansubject[1][]{%
  \tikz[overlay, remember picture]{
    \node[inner sep=0pt, anchor=text] (humansubjectnode) {\phantom{human-subject}};
    \draw[color3, line width=2pt] 
      ([yshift=-1pt]humansubjectnode.south west) --
      ([yshift=-1pt]humansubjectnode.south east);
  }\mbox{human\mbox{-}subject}\xspace%
}

\newcommand\humansubjects[1][]{
  \tikz[overlay, remember picture]{
    \node[inner sep=0pt, anchor=text] (humansubjectsnode) {\phantom{human subjects}};
    \draw[color3,  line width=2pt] 
      ([yshift=-1pt]humansubjectsnode.south west) --
      ([yshift=-1pt]humansubjectsnode.south east);
  }\mbox{human~subjects}\xspace%
}

\newcommand\Humansubject[1][]{%
  \tikz[overlay, remember picture]{
    \node[inner sep=0pt, anchor=text] (Humansubjectnode) {\phantom{Human-subject}};
    \draw[color3, line width=2pt] 
      ([yshift=-1pt]Humansubjectnode.south west) --
      ([yshift=-1pt]Humansubjectnode.south east);
  }\mbox{Human\mbox{-}subject}\xspace%
}

\newcommand\humanparticipant[1][]{%
  \tikz[overlay, remember picture]{
    \node[inner sep=0pt, anchor=text] (humanparticipantnode) {\phantom{human~participant}};
    \draw[color3, line width=2pt]
      ([yshift=-1pt]humanparticipantnode.south west) --
      ([yshift=-1pt]humanparticipantnode.south east);
  }\mbox{human~participant}\xspace%
}

\newcommand\humanparticipants[1][]{%
  \tikz[overlay, remember picture]{
    \node[inner sep=0pt, anchor=text] (humanparticipantsnode) {\phantom{human~participants}};
    \draw[color3,  line width=2pt] 
      ([yshift=-1pt]humanparticipantsnode.south west) --
      ([yshift=-1pt]humanparticipantsnode.south east);
  }\mbox{human~participants}\xspace%
}

\newcommand\Humanparticipants[1][]{
  \tikz[overlay, remember picture]{
    \node[inner sep=0pt, anchor=text] (Humanparticipantsnode) {\phantom{Human~participants}};
    \draw[color3,  line width=2pt] 
      ([yshift=-1pt]Humanparticipantsnode.south west) --
      ([yshift=-1pt]Humanparticipantsnode.south east);
  }\mbox{Human~participants}\xspace%
}

\usepackage{xspace}
\newcommand{\untrained}{\textbf{Untrained}\xspace}
\newcommand{\trained}{\textbf{Trained}\xspace}
\newcommand{\expert}{\textbf{Expert}\xspace}
\newcommand{\experts}{\textbf{Experts}\xspace}
\newcommand{\tuned}{\textbf{Tuned}\xspace}

\crefname{section}{Section}{Sections}
\crefname{figure}{Fig.}{Figs.}
\crefname{table}{Table}{Tables}

\begin{document}

\title{Exploring MLLMs Perception of Network Visualization Principles}
\author{%
  Jacob Miller*,
  Markus Wallinger*,
  Ludwig Felder,
  Timo Brand,\\
  Henry Förster,
  Johannes Zink,
  Chunyang Chen,
  Stephen Kobourov
  \thanks{
   * denotes equal contribution.
  	All authors are with the Technical University of Munich.
  	E-mail: firstname.lastname@tum.de.
}
}

\maketitle

\begin{abstract}
 In this paper, we test whether Multimodal Large Language Models (MLLMs) can match human-subject performance in tasks involving the perception of properties in network layouts. Specifically, we replicate a human-subject experiment about perceiving quality (namely stress) in network layouts using GPT-4o,  Gemini-2.5 and Qwen2.5. 
 Our experiments show that giving MLLMs the same study information as trained human participants yields performance comparable to that of human experts and exceeds that of untrained non-experts. 
 Additionally, we show that prompt engineering that deviates from the human-subject experiment can lead to better-than-human performance in some settings. 
 Interestingly, like human subjects, the MLLMs seem to rely on visual proxies rather than computing the actual value of stress, indicating some sense or facsimile of perception. 
 Explanations from the models are similar to those used by the human participants (e.g., an even distribution of nodes and uniform edge lengths). 
\end{abstract}


\begin{figure*}
  \centering
  \includegraphics[alt={ChatGPT and Gemini take part in an experiment on the perception of stress in network visualization.},page=3,width=\linewidth]{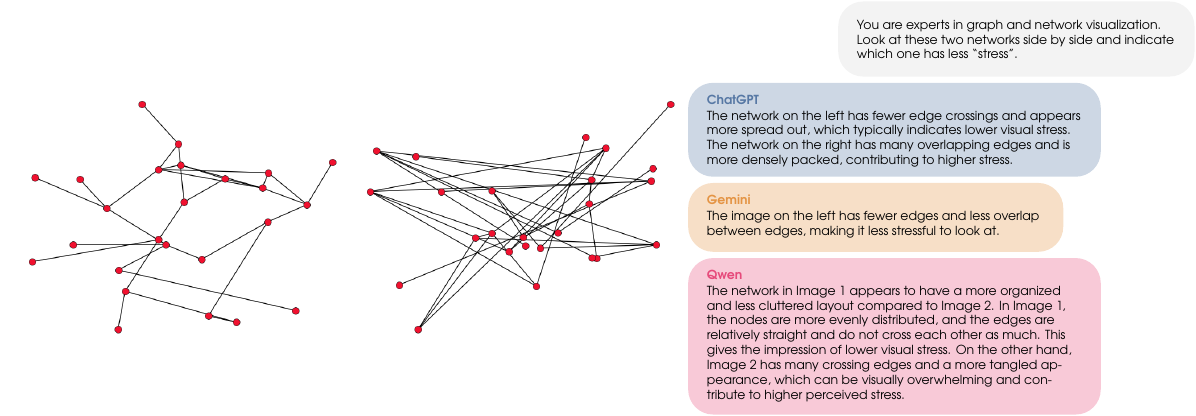}
  \caption{%
  	An illustrative example from our MLLM experiment. Left: a pair of different network diagrams of the same network is shown to an MLLM. Right: the gray box summarizes the prompt, asking you to identify the diagram with lower stress. The blue, orange, and red boxes show parts of the responses from ChatGPT, Gemini, {and Qwen}, respectively. While all MLLMs answered correctly, the explanation from Gemini contains incorrect parts (the number of edges is the same in both diagrams, as they represent the same network).
  }
  \label{fig:teaser}
\end{figure*}

\section{Introduction}
\IEEEPARstart{N}{etwork} data represents complex real-world processes, but there can be many different visualizations of the same underlying data. 
Even for node-link diagrams alone, there are many possible network features one might want to emphasize, and many techniques for producing them have been proposed. For example, one can highlight local~\cite{DBLP:journals/cgf/KruigerRMKKT17,DBLP:journals/tvcg/ZhuCHHLZ21} or global~\cite{DBLP:conf/gd/GansnerKN04} structures, clusters~\cite{DBLP:journals/jgaa/Noack07}, or even particular network properties of interest~\cite{DBLP:conf/gd/ForsterKDE0KLMM24}. 
Evaluating or comparing techniques using established quality metrics such as \textit{stress} is common. 
Informally, stress measures how well the network layout captures the shortest-path distances between pairs of nodes. 
Lower stress indicates a better representation of the network topology. Judging stress visually involves several perceptual processes such as estimating length, distribution, and symmetry. Recently, an empirical study~\cite{DBLP:conf/gd/MooneyPWK024} determined that human experts can reliably decide which of two diagrams has lower stress and that non-experts can be trained to do so, too.

Multimodal Large Language Models (MLLMs) are AI systems capable of understanding and generating content across multiple data types, such as text, images, audio, and video.
They have rapidly become ubiquitous across all fields of science and excel at producing rapid results to general inquiries via text or image prompts. 
Several high-quality models have recently become available to the public, such as OpenAI's GPT-4o\footnote{\url{https://platform.openai.com/docs/models/gpt-4o}}, Google's Gemini-2.5\footnote{\url{https://ai.google.dev/models}}, and Alibaba Cloud's Qwen2.5\footnote{\url{https://www.alibabacloud.com/help/en/model-studio/what-is-qwen-llm}}.
These models seem to achieve a high level of understanding of several topics, capable of matching or exceeding human-subject performance in the SAT and LSAT exams, as well as competitive programming challenges~\cite{Kishky25}. 

With advances in these models' vision capabilities, a natural question is what exactly their capabilities and limitations are for understanding visualizations? %
Recent experiments testing the general visualization understanding of ChatGPT-4 and Gemini, using the Visualization Literacy Assessment Test~\cite{DBLP:journals/tvcg/BendeckS25,hong2025llms}, revealed that both models can handle tasks such as comparing scatterplots or finding correlations fairly well. However, neither model performed consistently well on other tasks, such as reading pie charts or histograms.  

What remains unclear is whether these models have a sense of perception that enables them to understand and interpret other complex visualizations, such as node-link diagrams. 
{ As visually judging stress involves several perceptual processes and some understanding of the network structure, it makes for a good candidate study to test MLLM understanding. Additionally, the results of Mooney et al.~\cite{DBLP:conf/gd/MooneyPWK024} can serve as a human-subject baseline for comparing MLLMs. }

The research landscape is rapidly evolving regarding MLLMs, and it is imperative that we, as a community, broaden our collective understanding of how these models might be used to draw conclusions about visualizations and how valid those conclusions might be. MLLMs show promise for visualization designers and researchers by potentially enabling thorough, inexpensive piloting for human-subject studies if they indeed possess human-like visualization literacy. 
{
MLLMs present a potential risk in the validity of future crowd-sourced studies as well, so understanding their current limitations and when they can pass as a human being is timely for the visualization community at large. 
{
A secondary motivation is that quantitatively evaluating a node-link diagram from an image (without the underlying data) is a task often asked of human-subject study participants, but this can be expensive in terms of cost and effort. 
If this task can be delegated to an MLLM with high reliability, it may enable smaller-scale studies with results of similar soundness.
}
}
We aim to begin to address these questions in the context of network visualization, a yet unexplored topic at the time of writing.

This paper evaluates how MLLMs interpret, perceive, and respond to the most common network visualizations -- node-link diagrams.
Specifically, we compare the performance of \chatgpt %
, \gemini, and \qwen ~by replicating a \humansubject ~experiment of Mooney et al.~\cite{DBLP:conf/gd/MooneyPWK024} where participants were given two images of node-link diagrams.
The participants had to decide whether the left or right image showed a node-link diagram with lower stress, or whether both had similar stress.
Our results show that, given the same training information, MLLMs mimic the performance of all three types of human participants (untrained novices, trained novices, and experts). %
 By fine-tuning the prompts, we obtain better performance with interesting deviations.
We extract and analyze the main topics from the LLMs' reasoning
and investigate correlations with the underlying properties of the network layouts.
The reasoning of the MLLMs suggests that, like human participants, they rely on visual cues when judging stress. 
In summary, our contributions are as follows.
\begin{itemize}
    \item We replicate, to our knowledge, for the first time, a network visualization human-subject study with MLLM participants.
    \item We statistically analyze MLLM performance compared to human-subject performance.
    \item We analyze how deviating from instructions given to human-subjects affects model performance.
    \item We investigate the reasoning behind the models' decisions.
\end{itemize}

All prompts, code, data, results, and analysis are provided in an open-source OSF repository: \url{https://osf.io/748mx/}. 

\section{Related Work and Background}
In Section~\ref{sec:related}, we discuss recent related work on the use of large language models (LLMs) in the visualization field, focusing on their use for visualization generation, recommendation, evaluation, and as study participants. In Section~\ref{sec:background}, we provide background information about network visualization, stress, and the perception-of-stress human-subjects experiment.

\subsection{Related Work}
\label{sec:related}

\noindent
\textbf{Generation and Recommendation:}
Text-based LLMs have had some success in applying visualization principles to produce plots and diagrams. 
Di Bartolomeo et al.~\cite{DBLP:conf/vissym/BartolomeoSSD23} teach an LLM to apply a simple layered drawing algorithm given input data (preventing the model from executing code).
Vazquez~\cite{DBLP:conf/pacificvis/Vazquez24} shows mixed results when using LLMs to create simple visualizations.
Using an LLM, Liew and Mueller~\cite{LiewM22} create captions for visualizations,
while Tian et al.~\cite{DBLP:journals/tvcg/TianCDYYZW25} generate charts.
There has also been recent work %
on understanding the use of LLMs and MLLMs for visualization design.
Wang et al.~\cite{DBLP:conf/emnlp/WangZWLW23} use LLMs to recommend the best visualization idiom for a given input dataset.
Podo et al.~\cite{DBLP:journals/corr/abs-2402-02167} propose an evaluation stack for LLM-generated visualizations.
Masry et al.~\cite{MasryTBKHJ25} are among the first to employ an MLLM
for understanding and reasoning about charts given as images.
They follow an instruction-tuning approach of a pre-trained MLLM.
{ Chen et al.~\cite{chen2024viseval} propose a visualization benchmark dataset to evaluate how (M)LLMs perform on generating visualizations from text.}

\noindent
\textbf{Evaluation:}
With the rise of multimodal LLMs, the question of whether (M)LLMs possess visualization literacy has received considerable attention.
This question has recently been investigated by
Li et al.~\cite{DBLP:journals/corr/abs-2407-10996},
Bendeck and Stasko~\cite{DBLP:journals/tvcg/BendeckS25},
and Hong et al.~\cite{hong2025llms}.
Previously, Chen et al.~\cite{ChenZWTWBGP23}
studied this question with a pure text-based LLM. %
Lo et al.~\cite{DBLP:journals/tvcg/LoQ25} use LLMs to identify misleading visualizations. 
Schetinger et al.~\cite{DBLP:journals/cgf/SchetingerBEMMPA23} analyze how generative models hallucinate or misrepresent visualizations given input data, despite their nice aesthetics.

Li et al.~\cite{DBLP:journals/corr/abs-2407-10996},
Bendeck and Stasko~\cite{DBLP:journals/tvcg/BendeckS25}
and Hong et al.~\cite{hong2025llms} evaluate the visualization literacy of MLLMs by replicating the visualization literacy assessment test (VLAT)~\cite{lee_vlat_2017}.
This test is an online study to assess 
performance on low-level tasks with common visualization idioms, e.g.,  bar charts, line charts,  scatterplots.
All three conduct the VLAT using an MLLM agent, providing the relevant visualization as an image and prompting the agent to answer a multiple-choice question.
Bendeck et al.\ and Hong et al.\ restrict the output of the model to a single letter, and disallow the ``Omit'' option, which is available 
in the standard VLAT (and where ``Omit" results in less penalty than an incorrect answer).

Our experiment differs at a high level, as our research questions primarily focus on understanding MLLM's performance in interpreting complex visualizations at a perceptual level.
In our experiments, we try to replicate as closely as possible a human-subjects experiment as opposed to Hong et al.\ who make several concessions (e.g.,  removing the ``Omit'' option from the VLAT). Additionally, we compare directly to the individual results of a recent human-subjects study, allowing us to conduct a statistical analysis unlike the previous approaches
where the comparison is limited to the aggregate accuracy of human subjects.

Wang et al.~\cite{DBLP:journals/tvcg/WangHTBB25} revisit a human-subject study in a visualization context.
They partially re-implement a case study
of Xiong et al.~\cite{DBLP:journals/tvcg/XiongSBKLF22} about predicting the main takeaway message of a bar chart. 
They provide
in-context examples,
which improves the performance of the MLLM, a phenomenon also observed for other tasks~\cite{NEURIPS2020_1457c0d6,DBLP:journals/tvcg/WangHTBB25}.
Wang et al.\ conclude mixed results and state:
``We hope our work can motivate investigation into other visualization types and additional dimensions of perceptual awareness.''
Instead of replicating a qualitative study with bar charts, we replicate a quantitative study with node-link diagrams to address these additional research questions.

\noindent
\textbf{MLLMs as Participants \& Users:}
Machine learning, and in particular LLMs and MLLMs, has become a topic of interest in visualization and human-computer interaction (HCI) research.
In recent years, several visualization conferences have hosted workshops on
``Machine Learning from User Interaction for Visualization and Analytics'' (at VIS),
``Exploring Research Opportunities for Natural Language, Text, and Data Visualization'' (at VIS),
``Visualization for AI Explainability'' (at VIS),
``Machine Learning Methods in Visualization for Big Data'' (at EuroVis),
and ``Visualization Meets AI'' (at PacificVis).

At the Conference on Human Factors in Computing Systems (CHI) 2024, %
two separate workshops discussed opportunities and risks associated with involving LLMs in different stages of research~\cite{DBLP:conf/chi/QuereSRGEPMBL24,DBLP:conf/chi/PrpaTWC24}.
A prior special interest group considered similar topics in a more focused context of computational social science~\cite{DBLP:conf/cscw/ShenLLPY23}. 
Pang et al.~\cite{CHI25} recently reviewed how LLMs have been applied in papers that appear in CHI from the years 2020 to 2025 and identified the usage of LLMs as participants and users as an emerging methodology in HCI research. H{\"{a}}m{\"{a}}l{\"{a}}inen et al.~\cite{DBLP:conf/chi/HamalainenTK23} provided a case study for this usage of LLMs in HCI contexts, and later Duan et al.~\cite{DBLP:conf/chi/DuanW0H24}, and Xiang et al.~\cite{DBLP:conf/chi/XiangZLCPJCS24} used LLMs to generate usability feedback on user interfaces.

In the context of data visualization, Hong et al.~\cite{hong2025llms} also mention the perspective of employing MLLMs as 
affordable and easily accessible substitutes for human subjects in evaluation studies. On the other hand, Agnew et al.~\cite{DBLP:conf/chi/AgnewBCDEPMM24} and H{\"{a}}m{\"{a}}l{\"{a}}inen et al.~\cite{DBLP:conf/chi/HamalainenTK23} discuss ethical concerns for such a usage of LLM systems. Another technical limitation is that LLMs can %
outperform human study participants; see, e.g., a %
study by Ziems et al.~\cite{DBLP:journals/coling/ZiemsHSCZY24}.

\subsection{Background}
\label{sec:background}

Networks naturally arise as mathematical models of relational systems, such as biological protein interactions~\cite{gao2023hierarchical}, power and infrastructure~\cite{DBLP:journals/nms/Dijck21}, and social interactions~\cite{pascual2022social}. Effective network visualization
is a core visualization topic.%

By far the most popular choice of idiom for network visualizations is the so-called \textit{node-link diagram}. In this setting, each entity in the network (called a node or a vertex) is encoded as a mark, such as a circle, while the presence of a relationship (called a link or an edge) between two nodes is encoded as a line segment between them. Evaluation of the effectiveness of node-link diagrams has historically been grounded in perceptual principles~\cite{DBLP:journals/vlc/Purchase02} with studies showing the aesthetic metrics of edge crossings, angular resolution, node uniformity, edge length uniformity, and several others to affect readability and task performance~\cite{DBLP:journals/ese/PurchaseCA02}. 

As there are many possible features of a network that one might want to emphasize in a node-link diagram, many techniques to produce them have been proposed. 
Some techniques highlight local structures~\cite{DBLP:journals/cgf/KruigerRMKKT17,DBLP:journals/tvcg/ZhuCHHLZ21}, by preserving small neighborhoods around nodes and possibly missing the large-scale structure.
Other techniques highlight global structures~\cite{DBLP:conf/gd/GansnerKN04}, attempting to capture a network's large-scale structure while allowing local errors. %
Techniques can also highlight clusters by forcing separation between groups of nodes~\cite{DBLP:journals/jgaa/Noack07}, or 
to ``visually prove'' specific network properties, such as the existence of a bridge in the network~\cite{DBLP:conf/gd/ForsterKDE0KLMM24}.
It is common to evaluate or compare techniques using established quality (faithfulness) metrics~\cite{DBLP:conf/apvis/NguyenEH13}, which express the distortion of a network property represented in the diagram. The most common quality metric of this type is \textit{stress}, the third most reported quantitative evaluation metric in network visualization (behind running time and the number of crossings)~\cite{DBLP:journals/cgf/BartolomeoCSPWD24}.

Stress is a family of quality metrics that measures the discrepancy between the graph-theoretic distances between nodes in the data (i.e., the number of edges on a shortest path between them), and the realized geometric distances between their coordinates in the diagram. Most commonly used is  ``normalized stress",  defined by Gansner et al.~\cite{DBLP:conf/gd/GansnerKN04}. 
The concept of stress dates back to the works of Torgerson~\cite{torgerson1952multidimensional}, Kruskal~\cite{kruskal1964multidimensional}, Shepard~\cite{shepard1962analysis}, and later Sammon~\cite{DBLP:journals/tc/Sammon69}, who initially utilized it as a statistical analysis tool. It was first used in network visualization as an optimization function proposed by Kamada and Kawai~\cite{DBLP:journals/ipl/KamadaK89}. Their technique was later improved by Gansner et al.~\cite{DBLP:conf/gd/GansnerKN04}, and Zheng et al.~\cite{DBLP:journals/tvcg/ZhengPG19} show that stochastic gradient descent is even more effective at optimizing stress. There are many variants of stress used as optimization functions~\cite{DBLP:journals/tvcg/GansnerHN13,DBLP:conf/gd/MillerHK23}, and many more examples of works which use stress as an evaluation metric despite not directly optimizing it~\cite{DBLP:journals/cgf/ArleoMA22,DBLP:conf/gd/HongETWCHLC19,DBLP:conf/gd/MarnerSTKEH14,DBLP:conf/gd/SimonettoAK17,DBLP:journals/cgf/WageningenMT24}. 

Stress has been shown to be a good proxy for symmetry~\cite{DBLP:journals/cgf/WelchK17}, and it has been shown that people tend to prefer diagrams with lower stress~\cite{chimani2014people}. 
A recent study of Mooney et al.\ attempts to answer the broad question, ``Can people see stress?~\cite{DBLP:conf/gd/MooneyPWK024}'' 
The study is further detailed in \cref{sec:human-exp}, but a key takeaway is that participants could reliably identify diagrams with lower stress, although the feedback and subsequent interviews indicated that they made this judgment at a perceptual level. In other words, although participants were asked directly about stress, they relied on high-level visual proxies to identify the stress of a diagram. In this work, we primarily ask whether these results extend to MLLMs and, if so, to what extent. If asked directly about stress, which has a well-defined mathematical definition, will an MLLM also engage in a similar high-level perceptual processing as human participants indicated?

\section{Experiment}

In this section, we first summarize the experiment by Mooney et al.~\cite{DBLP:conf/gd/MooneyPWK024} before we introduce our research questions.

\subsection{Reference experiment}
\label{sec:human-exp}

We now describe the study design of Mooney et al.~\cite{DBLP:conf/gd/MooneyPWK024}, on which our experiment is based. They investigate whether participants can ``see'' stress using a paired-stimulus experiment: participants are shown a pair of diagrams of the same network and asked to choose the one with lower stress. 

The controlled variable in the experiment is the difference in stress between the two diagrams. In particular, the authors use the Kruskal stress metric (KSM)~\cite{kruskal1964multidimensional}, also known as non-metric stress. KSM is bound to a $[0,1]$ range, is scale-invariant, and 
 is formally defined as follows: 
\[\sqrt{\frac{\sum_{i,j} \left( ||X_i - X_j|| - \hat{d}_{i,j}   \right)^2}{\sum_{i,j} ||X_i - X_j||^2}}\]
where the sum is over all pairs $i, j$ of nodes in the network,
$X_i$ is the coordinate position of node $i$, and $\hat{d}_{i,j}$
is a notion of distance between nodes $i$ and $j$ found from a monotonic regression on the Shepard diagram~\cite{shepard1962analysis} of the drawing.

The study contained three groups of participants: \trained novices, \untrained novices, and \experts. The instructions and the training for the different groups varied as follows. \trained novices were first introduced to concepts of network visualization (nodes, edges, diagrams, etc.) and were presented with the concept and definition of stress. Next came a round of 9 training questions, and after each answer, participants received feedback (correct or incorrect). Finally, the participants answered 45 questions. \untrained participants received the same introduction but were not given any feedback on the training questions. The \experts (identified by the authors) were given just the two-sentence instruction to identify which of the two drawings had lower stress.
Participants in all three settings were shown, in random order, 45 pairs of network diagrams and were asked to identify the diagram with lower stress, with choices of ``The drawing on the left has lower stress'', ``The drawings have the same stress'', and ``The drawing on the right has lower stress.'' After each response, participants were asked to report their confidence (either ``confident'' or ``not confident''). For each network size considered in the study (10, 25, and 50 nodes), there were five unique networks with many diagrams at KSM levels between 0.4 and 0.8, with intervals of 0.05. This gives nine unique KSM level differences for each of the five networks. Finally, 
the participants were asked about ``the overall strategy used to determine which drawing had lower stress'' and 
some demographic and free-response feedback was collected. 

The results showed that all participant groups could reliably identify the network diagram with lower stress in most cases, with overall accuracy being 68\%, 75\%, and 78\% for the \untrained, \trained, and \expert groups, respectively, compared to the baseline of 33\% for guessing.
Notably, training helped to improve the performance:
the difference in mean accuracy between \untrained and \trained participants was statistically significant.

However, comments in response to the question about ``the overall strategy used to determine which drawing had lower stress'' indicated that participants made their decisions based on what ``looked right'' or ``felt right''. 
More detailed responses mentioned features specific to aesthetic criteria, such as node distribution, edge crossings, and edge length uniformity.
This indicates that participants were not necessarily performing low-level tasks on the data visualization but making a high-level perceptual judgment of the aesthetic appeal of the two diagrams via visual proxies.

Since human participants could generally differentiate between low- and high-stress diagrams
without actually computing the stress,
it is a natural question to ask how MLLMs perform in the same test.
Can they reach similar accuracy levels?
In about a quarter of the instances, human participant judgments were wrong~--
do MLLMs possess or gain a better notion of salient differences in low- and high-stress diagrams?
Can MLLMs potentially exceed human participants in terms of accuracy?
In the context of explaining answers given by MLLMs, we are also interested in seeing if MLLMs resort to visual proxies or perceptual-level judgments
as humans did.

\subsection{Research Questions}

Our experiment tries to shine a light on these research questions: 

\begin{itemize}
    \item \textbf{RQ1}: Given the same information as \humanparticipants, how effectively do \chatgpt, \gemini, and \qwen\ identify lower-stress node-link diagram compared to human performance?
    \item { \textbf{RQ2}: Can MLLMs achieve greater-than-human accuracy in identifying lower-stress node-link diagrams?}
    \item \textbf{RQ3}: Is there a relationship between the properties of diagrams and the given answers by the MLLMs?  
    \item \textbf{RQ4}: Can we obtain any insight into the decision-making process of the MLLMs?
\end{itemize}

In \textbf{RQ1}, we focus on replicating the experiment by Mooney et al.~\cite{DBLP:conf/gd/MooneyPWK024}. 
However, instead of \humanparticipants, \chatgpt, \gemini, and \qwen\ perform the study. This entails giving the same relevant study information and training examples as prompts to the MLLMs. We then focus on three aspects: firstly, we investigate how the performance differs between the models. Secondly, as different levels of training were used in the original experiment~\cite{DBLP:conf/gd/MooneyPWK024}, we investigate how this affects the models' performance when given the same information. Thirdly, we compare the performance of the MLLMs to the performance of the \humanparticipants.

{ In \textbf{RQ2}, we answer if it is possible to increase the performance of the MLLMs when they are presented
with more appropriate information than in \textbf{RQ1}. }

For \textbf{RQ3}, we perform a statistical analysis between the properties of the diagrams and the given answers. 
Lastly, in \textbf{RQ4}, we shift over to the reasoning the models provide, together with their answers. We investigate what visual cues the models use to infer an answer. Also, we try to identify similarities between the models' answers and those provided by \humanparticipants.

\section{Experimental Setup}

Here we describe our experimental design to address the above research questions. All of our prompts, data collection scripts, and final results are made available as supplemental material online.

{
Although there are many MLLMs available, we chose to use OpenAI's \chatgpt (gpt-4o-2024-08-06) and Google's \gemini (gemini-2.5-pro-exp-03-25) for their wide popularity and high quality~\cite{chen2024we, lu2024wildvision}, lending our experiment external validity at the expense of using closed-source models.
To better ensure replicability, we also use an open-source model, Alibaba Cloud's \qwen (vl-72b-instruct)~\cite{bai2025qwen25vltechnicalreport}, which, at the time of writing (June 2025), is highly rated for its vision capabilities.
}
{ 
The test is conducted using API requests, either through the proprietary interfaces for \chatgpt and \gemini, or via \ url {https://openrouter.ai} for \qwen. 
Each request starts a new model session, so the model cannot ``remember'' the stimuli for future trials or experiments.
}

\paragraph*{\textbf{Stimuli}}
The networks used for the stimuli are the same as in the original study~\cite{DBLP:conf/gd/MooneyPWK024}, 
i.e., Erdős–Rény random networks with 10, 25, and 50 vertices. 
The same procedure for generating node-link diagrams with different stress values was used, but the instances differ from those used in the original study, in case either MLLM may have seen them during its training.
For each setting of trained, untrained, and expert, 
the MLLMs are independently asked the study question on 216 pairs of networks.
These were selected by sampling for every stress level difference, 
i.e., every value between 0.4 and 0.8 with 0.05 increments.
{
One corpus of drawings was used for initial testing, and a second, separate corpus was used for the final experiment.
}

\subsection{Replicating the Study} \label{sec:prompts}
To address \textbf{RQ1}, it is important to carefully give the model the \textit{exact} information a \humanparticipant\ would have received. This means that, like in the reference experiment, we should have three prompt conditions. 
We detail how we presented this information to the models, from most to least context provided.

\begin{figure}[t]
\begin{mdframed}[roundcorner=10pt]
    \centering
    \includegraphics[width=0.99\linewidth, trim={2.2cm 22.25cm 2.2cm 2.2cm},clip]{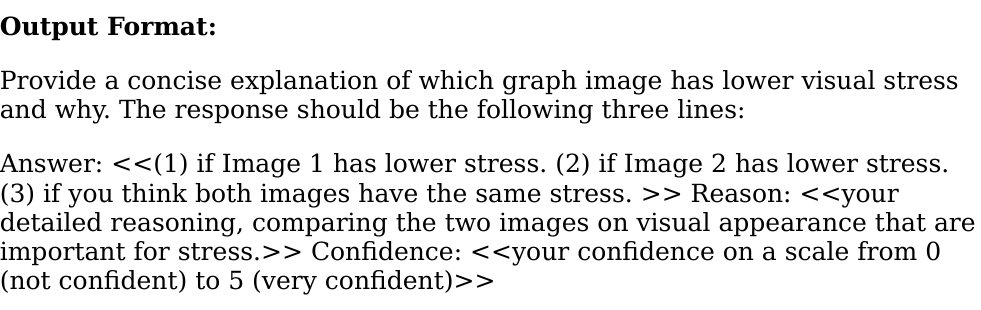}    
\end{mdframed}
    \caption{Specific instructions to the MLLMs on structuring the response to stimuli. The instructions are given in Markdown.}
    \label{fig:prompt-onetwo}
\end{figure}

\noindent
\trained participants in the \humansubject ~study were given the most amount of information and feedback on their results. This included the following three parts:
\begin{itemize}
    \item an explanation of what the study will ask,
    \item illustrated definitions of network concepts and stress, and
    \item nine training examples, where participants were told if their response was correct or incorrect.
\end{itemize}
The first point (the study explanation) closely matches that of Mooney et al.~\cite{DBLP:conf/gd/MooneyPWK024}. We removed identifiable contact information of the study conductors from the introduction, but we pre-pended ``You are a person participating in a study.'' to the beginning of the explanation, and instructed the model on how to respond; see Fig.~\ref{fig:prompt-onetwo}.  
This instruction is given to the model as a system prompt, pre-pended to each API request, which guides its responses to subsequent user queries.

The network definitions and illustrations are unchanged from the original study.
Finally, the nine training examples are shown as pairs, one after another.
Because the model does not ``remember'' a previous API request, it does not make sense to only show the correct answer after a response.
Therefore, we give the correct option along with all nine training examples. 
These two points are sent in user mode, which is repeated for every trial; the model is ``reminded'' of the training sequence since it does not ``remember'' the previous trials.

After these instructions, the model is shown a pair of network diagrams, each presented as its own image with accompanying text, ``Image 1'' and ``Image 2'', respectively. The model is then prompted: ``Which image has lower stress?''
The \untrained\ \humanparticipants\ had less context provided to them in the Mooney et al.~\cite{DBLP:conf/gd/MooneyPWK024} study. In particular, they received 
the first two of the three bullets from the \trained setting.
Notably, feedback on the training examples was not provided to the participants (though they still completed them), and these results were ignored.

There are minor changes to the explanations from the \trained participants that we also update here; the \trained participant explanations refer to the training examples, while the \untrained explanations do not. In this mode, we remove the training examples entirely, leaving only the stress definitions with illustrations for context.

The \expert \humanparticipants\ received the least amount of context. In fact, they only received the first of the three bullets from the \trained setting (and this explanation was significantly reduced). 
\experts were told they would see two network diagrams and should indicate which had less stress -- it was expected they understood these definitions. 

A notable change in this mode in our study is that, instead of telling the MLLM, ``You are a person participating in a study,'' we add, ``You are an expert in graph and network visualization'' to the initial explanation to specify the desired MLLM persona. The illustrated definitions are completely removed in this setting.

\begin{figure*}[ht]
    \centering

    \parbox[c]{0.24\linewidth}{\centering \hspace{0.4cm} \textbf{Trained}}
    \parbox[c]{0.24\linewidth}{\centering \hspace{0.4cm}\textbf{Untrained}}
    \parbox[c]{0.24\linewidth}{\centering \hspace{0.4cm}\textbf{Expert}}
    \parbox[c]{0.24\linewidth}{\centering \hspace{0.4cm}\textbf{Tuned}}    

    \includegraphics[width=0.24\linewidth]{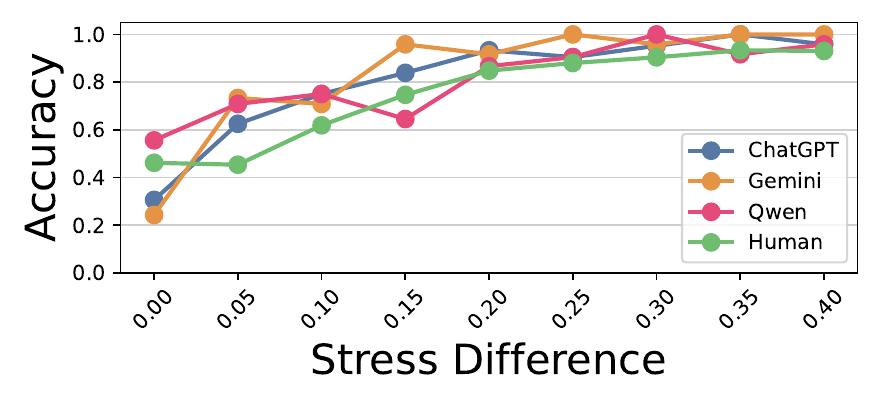}
    \includegraphics[width=0.24\linewidth]{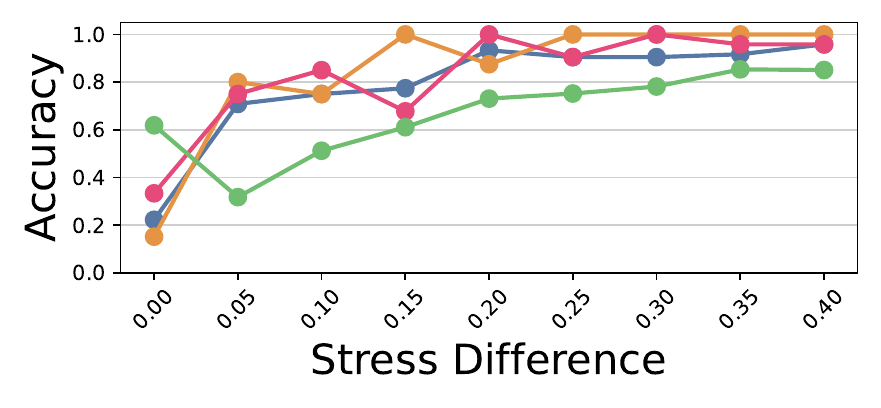}
    \includegraphics[width=0.24\linewidth]{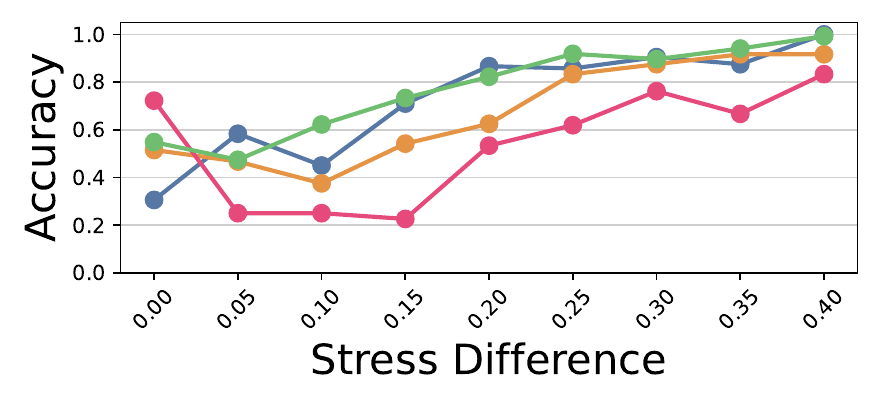}
    \includegraphics[width=0.24\linewidth]{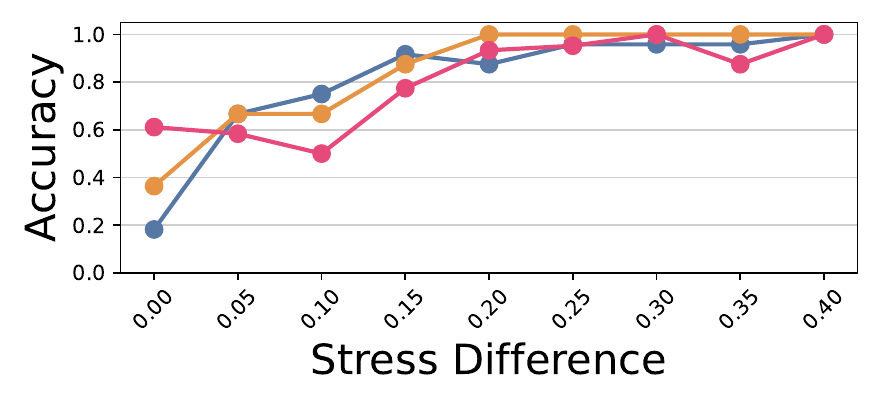}

    \caption{Overall accuracy in the \textbf{Trained}, \textbf{Untrained}, \textbf{Expert} and \textbf{Tuned} 
    setting with respect to the stress level difference. 
    }
    \label{fig:linechart}
\end{figure*}
\subsection{Tuning the Prompts}
\label{sec:prompt-engineering}

While the prompts from \cref{sec:prompts} focus on being as close as possible to the instructions from the \humansubject ~experiment, they might be suboptimal for an MLLM.
Hence, we focused on improving performance by tuning a prompt.
We show snippets from the prompt and explain the overall strategy.
The full prompt is available in the supplemental material.

{
Several strategies and recommendations for better prompts have been proposed~\cite{openai,reynolds2021prompt}, and recent surveys~\cite{schulhoff2024,Sahoo20204} catalog dozens of established techniques.
Among the most relevant to our setting are few-shot prompting~\cite{NEURIPS2020_1457c0d6}, where solved examples guide the model's responses;
chain-of-thought (CoT) prompting~\cite{Wei0SBIXCLZ22}, where complex reasoning is decomposed into intermediate steps; comparative prompting~\cite{gozzi2024comparative}, where the model is explicitly asked to compare and contrast two inputs; multimodal contextualization~\cite{DBLP:conf/eccv/DovehPMLAAUK24}, where non-text inputs are described to the model;
and role prompting~\cite{schulhoff2024}, where the model is assigned a specific persona or expertise level.
Other techniques, such as self-consistency~\cite{WangWSLCNCZ2023} and tree of thoughts~\cite{YaoYZS00N23}, have shown promise on complex reasoning benchmarks, but require multiple model calls per stimulus and are thus impractical in our setting, where each of the 216 trials per condition is already an independent API request.
We leave exploration of such multi-call strategies to future work.

}

{
We summarize the \tuned prompt with several components that realize the prompting strategies described above.
First, we provide a precise role~\cite{schulhoff2024} for the MLLM and describe the images it will receive. This implements role prompting by assigning the MLLM an expert persona and multimodal contextualization.
Second, we carefully explain stress and how it can be visually evaluated, followed by a detailed description of the evaluation criteria (node distribution, edge lengths, and crossings). 
This follows the chain-of-thought strategy~\cite{Wei0SBIXCLZ22}: rather than asking the model to judge stress directly, we decompose the task into structured evaluation criteria with clear section delimiters, guiding the model through intermediate reasoning steps.
Third, we provide more detail on how the evaluation process should work, including a step-by-step evaluation process and a typical recommendation~\cite{Wei0SBIXCLZ22} for splitting complex tasks into subtasks.
Finally, we provide several pairs of drawings as training examples, utilizing few-shot~\cite{NEURIPS2020_1457c0d6} and comparative~\cite{gozzi2024comparative} prompting strategies.
}

\section{Evaluation}
We report a summary of the replication experiment. 
The MLLMs perform well in the \trained setting, with \chatgpt\ achieving a $77.3\%$ overall accuracy, \gemini ~achieving $81.5\%$, and \qwen reaching $78.7\%$ accuracy. 
This is comparable to \humanparticipants, who reach $75.3\%$ when \trained. Despite numerically similar aggregate scores, we observe large differences in behavior with respect to the stress level, as seen in Fig.~\ref{fig:linechart} and \cref{tab:merged}. 
The MLLMs seem to struggle with the 0-level difference~-- where stress is the same. Human-subject accuracy is also lower in this condition, around 0.5, but is noticeably higher than that of the proprietary language models, which have near-0 accuracy (worse than random guessing). The big exception to this is \qwen in the \textbf{Expert} setting, which achieves $72\%$ accuracy for the 0-level stress difference, despite having poor accuracy overall. The accuracy of the MLLMs otherwise improves rapidly as the stress difference increases, with near-perfect accuracy for the larger networks.

In the {\untrained} setting, \chatgpt, \gemini, \qwen, and \humanparticipants ~achieve accuracy 
of $74.5\%$, $81.5\%$, $78.2\%$ and $67\%$.
In the {\expert} setting, the accuracies are $69.9\%$, $67.6\%$, $54.2\%$, and $77.2\%$.
We can see that there is little difference in the performance of the MLLMs for the \trained and \untrained settings.
With only the \expert instructions, performance worsens across all models.
Accuracies broken down by stress difference and network size are shown in Table~\ref{tab:merged}. Further details are available in the supplemental material.

{
\setlength{\tabcolsep}{5pt}
\begin{table*}[t]
\caption{Aggregated results in the \textbf{Trained}, \textbf{Untrained}, and \textbf{Expert}
setting for~\chatgpt,~\gemini,~\qwen, and~\human, and \textbf{Tuned} for the MLLMs.
The rows show the absolute stress level difference between the two network diagrams shown to the MLLM and participants. The columns show results for a combination of settings and models. Numbers colored in red are worse than chance ($\frac{1}{3}$), and underlined numbers show below-human performance (\textbf{Tuned} MLLMs are compared to \humansubject \experts).}
 \label{tab:merged}
\centering
\sffamily
\small
\renewcommand{\arraystretch}{1.25}
\sffamily
\tiny
\begin{tabular}{
c
c
c
c
c
>{\columncolor[HTML]{efefef}}c
>{\columncolor[HTML]{efefef}}c
>{\columncolor[HTML]{efefef}}c
>{\columncolor[HTML]{efefef}}c
c
c
c
c
>{\columncolor[HTML]{efefef}}c
>{\columncolor[HTML]{efefef}}c
>{\columncolor[HTML]{efefef}}c}
\rowcolor[HTML]{343434}
 & \multicolumn{4}{c}{\color[HTML]{F8F8F8}\textbf{Trained}} & \multicolumn{4}{c}{\color[HTML]{F8F8F8}\textbf{Untrained}} & \multicolumn{4}{c}{\color[HTML]{F8F8F8}\textbf{Expert}} & \multicolumn{3}{c}{\color[HTML]{F8F8F8}\textbf{Tuned}} \\
\rowcolor[HTML]{343434}
\multirow{-2}{*}{\color[HTML]{F8F8F8} \textbf{Stress Difference}}                                & \cellcolor{color1!60}\textbf{GPT-4o} & \cellcolor{color2!60}\textbf{Gemini-2.5} & \cellcolor{color4!60}\textbf{Qwen2.5} & \cellcolor{color3!60}\textbf{Human}  & \cellcolor{color1!60}\textbf{GPT-4o} & \cellcolor{color2!60}\textbf{Gemini-2.5} & \cellcolor{color4!60}\textbf{Qwen2.5} & \cellcolor{color3!60}\textbf{Human}  & \cellcolor{color1!60}\textbf{GPT-4o} & \cellcolor{color2!60}\textbf{Gemini-2.5} & \cellcolor{color4!60}\textbf{Qwen2.5} & \cellcolor{color3!60}\textbf{Human}  & \cellcolor{color1!60}\textbf{GPT-4o} & \cellcolor{color2!60}\textbf{Gemini-2.5} & \cellcolor{color4!60}\textbf{Qwen2.5} \\
\cellcolor[HTML]{343434}{\color[HTML]{F8F8F8} \textbf{0.00}} & \color{red}\underline{0.306} & \color{red}\underline{0.242} & 0.556 & 0.461 & \color{red}\underline{0.222} & \color{red}\underline{0.152} & \color{red}\underline{0.333} & 0.619 & \color{red}\underline{0.306} & \underline{0.515} & 0.722 & 0.548 & \color{red}\underline{0.182} & \underline{0.364} & 0.611 \\
\cellcolor[HTML]{343434}{\color[HTML]{F8F8F8} \textbf{0.05}} & 0.625 & 0.733 & 0.708 & 0.453 & 0.708 & 0.8 & 0.75 & \color{red}0.317 & 0.583 & \underline{0.467} & \underline{0.25} & 0.474 & 0.667 & 0.667 & 0.583 \\
\cellcolor[HTML]{343434}{\color[HTML]{F8F8F8} \textbf{0.10}} & 0.75 & 0.708 & 0.75 & 0.619 & 0.75 & 0.75 & 0.85 & 0.512 & \underline{0.45} & \underline{0.375} & \underline{0.25} & 0.622 & 0.75 & 0.667 & \underline{0.5} \\
\cellcolor[HTML]{343434}{\color[HTML]{F8F8F8} \textbf{0.15}} & 0.839 & 0.958 & \underline{0.645} & 0.747 & 0.774 & 1.0 & 0.677 & 0.611 & \underline{0.71} & \underline{0.542} & \underline{0.226} & 0.733 & 0.917 & 0.875 & 0.774 \\
\cellcolor[HTML]{343434}{\color[HTML]{F8F8F8} \textbf{0.20}} & 0.933 & 0.917 & 0.867 & 0.848 & 0.933 & 0.875 & 1.0 & 0.731 & 0.867 & \underline{0.625} & \underline{0.533} & 0.822 & 0.875 & {\phantom{0}1.0\phantom{0}} & 0.933 \\
\cellcolor[HTML]{343434}{\color[HTML]{F8F8F8} \textbf{0.25}} & 0.905 & {\phantom{0}1.0\phantom{0}} & 0.905 & 0.88 & 0.905 & 1.0 & 0.905 & 0.752 & \underline{0.857} & \underline{0.833} & \underline{0.619} & 0.919 & 0.958 & {\phantom{0}1.0\phantom{0}} & 0.952 \\
\cellcolor[HTML]{343434}{\color[HTML]{F8F8F8} \textbf{0.30}} & 0.952 & 0.958 & 1.0 & 0.904 & 0.905 & 1.0 & 1.0 & 0.781 & 0.905 & \underline{0.875} & \underline{0.762} & 0.896 & 0.958 & {\phantom{0}1.0\phantom{0}} & 1.0 \\
\cellcolor[HTML]{343434}{\color[HTML]{F8F8F8} \textbf{0.35}} & {\phantom{0}1.0\phantom{0}} & {\phantom{0}1.0\phantom{0}} & 0.917 & 0.933 & 0.917 & 1.0 & 0.958 & 0.853 & \underline{0.875} & \underline{0.917} & \underline{0.667} & 0.941 & 0.958 & {\phantom{0}1.0\phantom{0}} & \underline{0.875} \\
\cellcolor[HTML]{343434}{\color[HTML]{F8F8F8} \textbf{0.40}} & 0.958 & {\phantom{0}1.0\phantom{0}} & 0.958 & 0.931 & 0.958 & 1.0 & 0.958 & 0.851 & {\phantom{0}1.0\phantom{0}} & \underline{0.917} & \underline{0.833} & 0.993 & {\phantom{0}1.0\phantom{0}} & {\phantom{0}1.0\phantom{0}} & 1.0 \\
\cellcolor[HTML]{343434}{\color[HTML]{F8F8F8} \textbf{Column Mean}} & \cellcolor{color1!30}0.773 & \cellcolor{color2!30}0.815 & \cellcolor{color4!30}0.787 & \cellcolor{color3!30}0.753 & \cellcolor{color1!30}0.745 & \cellcolor{color2!30}0.815 & \cellcolor{color4!30}0.782 & \cellcolor{color3!30}0.67 & \cellcolor{color1!30}\underline{0.699} & \cellcolor{color2!30}\underline{0.676} & \cellcolor{color4!30}\underline{0.542} & \cellcolor{color3!30}0.772 & \cellcolor{color1!30}0.787 & \cellcolor{color2!30}0.829 & \cellcolor{color4!30}0.787 \\
\end{tabular}

\end{table*}
}

\subsection{RQ1}
After collecting results from the MLLMs, we aim to first compare their performance with human-subjects from the same experiment. 

\begin{quote}
\begin{mdframed}[roundcorner=10pt]
    {\bf RQ1:} Given the same information as \humanparticipants, how effectively do the MLLMs identify the lower-stress node-link diagram when compared to human performance?
\end{mdframed}
\end{quote}

The first hurdle to address in this comparison is the difference in data collection: while making requests to an MLLM results in independent samples (as each request is a different session), this is not true for \humanparticipants ~whose responses are affected by the order of stimuli and other factors.  

\begin{figure*}[ht]
    \centering
    \includegraphics[width=0.32\linewidth]{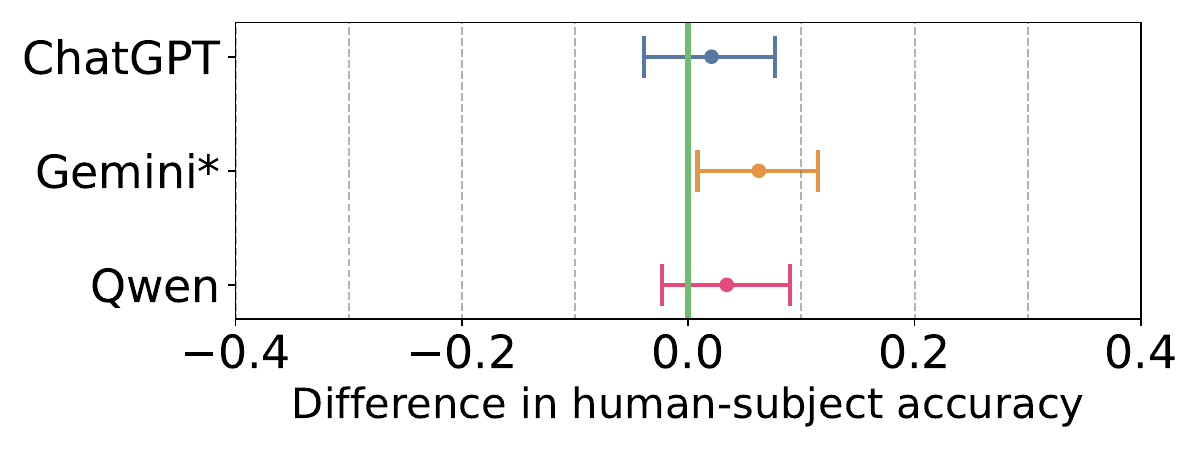}
    \includegraphics[width=0.32\linewidth]{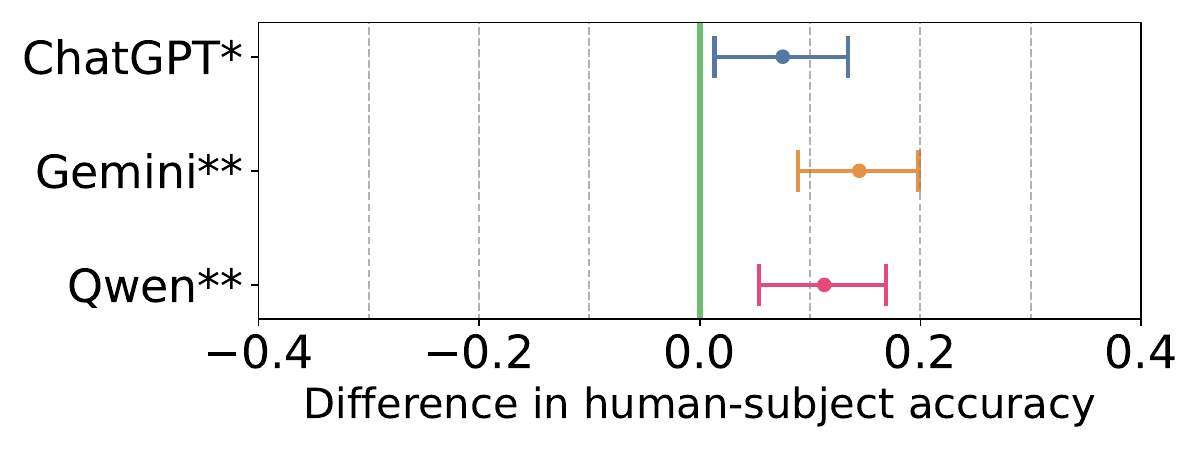}
    \includegraphics[width=0.32\linewidth]{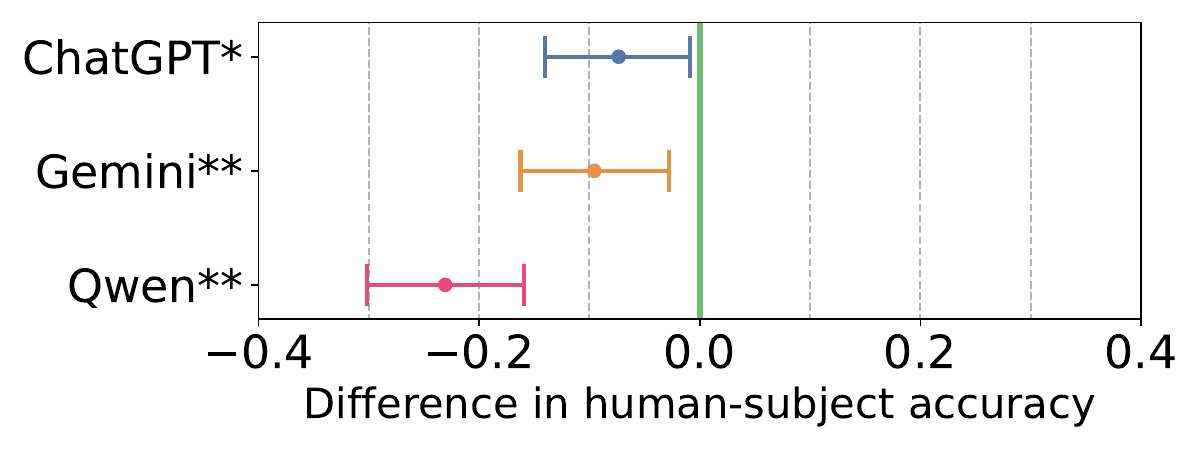}

    \includegraphics[width=0.32\linewidth]{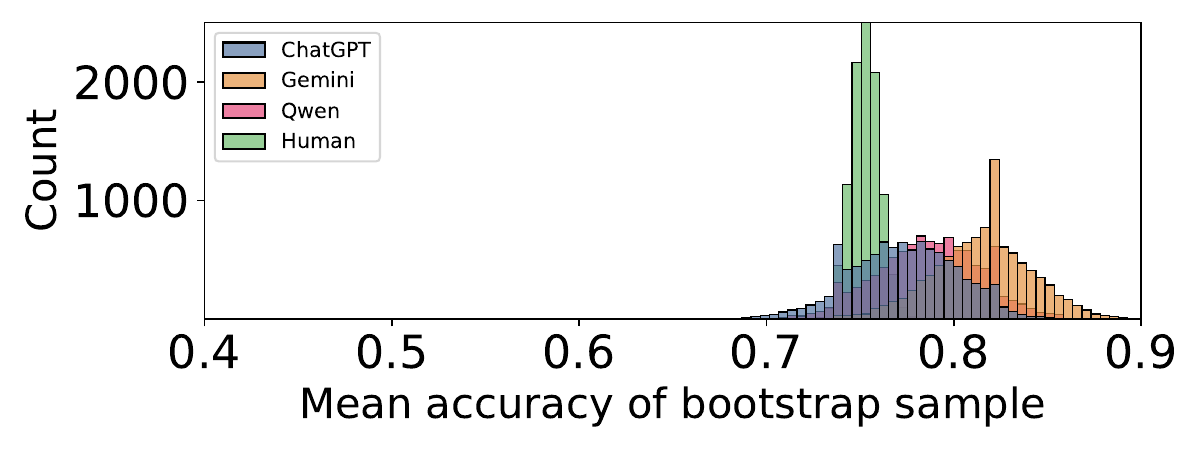}
    \includegraphics[width=0.32\linewidth]{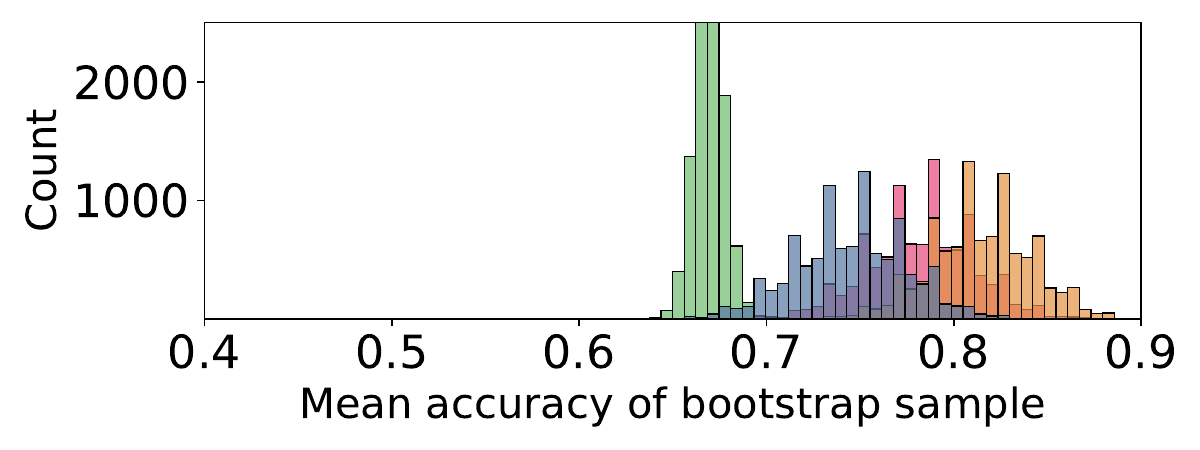}
    \includegraphics[width=0.32\linewidth]{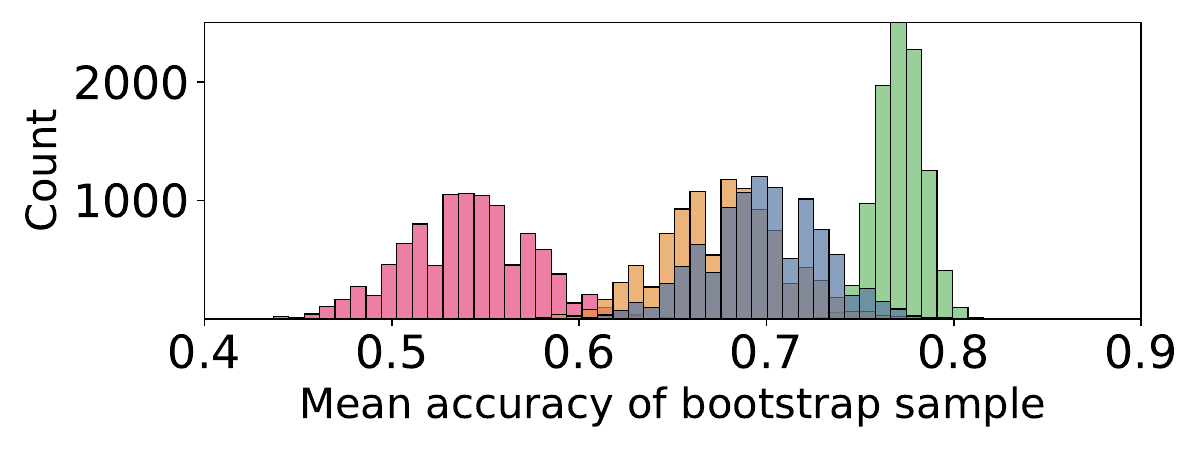}    

    \parbox[c]{0.32\linewidth}{\centering \hspace{0.4cm} (a) \trained}
    \parbox[c]{0.32\linewidth}{\centering \hspace{0.4cm} (b) \untrained}
    \parbox[c]{0.32\linewidth}{\centering \hspace{0.6cm} (c) \expert}

    \caption{(Top) 95\% confidence intervals for the 10,000 iteration bootstrapped difference of means test. 
    An interval containing zero indicates an absence of significant difference to \humansubjects. An interval entirely to the right (left) of zero indicates significantly better (worse) performance than \humansubjects. (Bottom) Histograms of the distributions of bootstrapped means for each setting. { A * indicates $p$-value $< 0.05$, ** indicates $p$-value $<0.01$.}}
    \label{fig:rq1_DoM}
\end{figure*}

We considered conducting traditional difference-of-means tests (i.e., $t$- or Wilcoxon tests), but this would require either comparing aggregate human-subject accuracy against individual MLLM responses or arbitrarily aggregating independent MLLM accuracies. The former is unfair to human responses, which are aggregated across several trials, and the latter introduces structure that is not present in the data. %

We instead employ a bootstrapping approach~\cite{tibshirani1993introduction}, which is a general method for statistical comparison and has been used in several visualization studies~\cite{brehmer2018visualizing,dragicevic2016fair,hong2025llms,tang2024comparative,vietinghoff2022detecting}.
The bootstrap statistical analysis method takes a data collection and creates many thousands of simulated samples (of the same size as the original) by drawing with replacement from the original data collection. For each simulated sample, one can compute statistics of interest, typically the mean or median, yielding a distribution of statistics. Given the large number of statistics, the resulting confidence intervals are reliable. The only assumption required of the original data is that it accurately represents the population from which it is sampled.

{
So, to analyze the difference between the MLLMs and \humansubject groups, we treat the individual responses as sample units, which results in 12 
total samples: three for each context (\trained, \untrained, \expert), and four for each agent (\chatgpt, \gemini, \qwen, \human). For all analyses, we perform 10,000 resamples. 
}

\paragraph*{Results}

{
To answer \textbf{RQ1}, we conduct a difference of means test between comparable groups -- \human ~with \chatgpt, \human with \gemini, and \human with \qwen. We do this for the \trained, \untrained, and \expert study settings.   
The confidence intervals of the test are shown in Fig.~\ref{fig:rq1_DoM}. 
In the \expert setting (least context),
human \expert participants outperform MLLMs (\chatgpt and \gemini) by about 7\%; this result is statistically significant in both cases. The \qwen model performs much worse overall; performing very well on the 0-level stress difference but poorly otherwise. Notably, the \expert participants were provided with the same amount of context as the MLLMs.  

In the \untrained setting (moderate context), we see the opposite result. Here, \chatgpt ~outperforms \untrained \humanparticipants ~by around 8\% while \gemini and \qwen ~outperform by nearly 15\%; these results are statistically significant. 
In the Mooney et al.~\cite{DBLP:conf/gd/MooneyPWK024} experiment, the \untrained participants were novices given some information but no training. This suggests that MLLMs can perform well with the information provided, and manual training is not as necessary as for human-subject participants.

Finally, in the \trained setting, participants were given the most %
context. We see that while \chatgpt and \qwen ~achieve around 3\% more accuracy than \trained \humanparticipants ~on average, the result is not significant. The \gemini model achieves 7\% higher accuracy, which is significant. %

\paragraph*{Discussion}
From our results, we draw several conclusions. Overall, MLLMs are competitive with \humanparticipants ~in identifying lower-stress network diagrams. Aside from the worst performing case for MLLMs (\expert \qwen), the lowest mean accuracy we see after bootstrapping is around 60\% (\expert \qwen). This is not so different from the lowest \humanparticipant ~mean accuracy of 67\% (\untrained). However, it is clear that the information, context, and prompt provided to an MLLM significantly affect accuracy. 
}

Specifically, the \expert setting for MLLMs performs relatively poorly, being more comparable to \untrained \humanparticipants. This could be for any number of reasons, for instance, the MLLMs may be relying on their prior knowledge that doesn't apply, such as the colloquial definition of stress. We investigate the MLLM reasoning in more detail in \cref{sec:RQ3}. Regardless, it is clear that MLLMs cannot perform as an \expert without more careful instruction.

We believe it is generally ill-advised to fully \mbox{replace} \human in visualization evaluation studies. %
The kinds of training and instructions given to an MLLM affect the performance in statistically significant ways. %
For instance, \trained MLLMs perform more similarly to \expert \humanparticipants than they do to \trained \humanparticipants. However, MLLMs show promise %
as helpful tools for pilot studies to generate quick feedback and check for floor/ceiling effects.%

\subsection{RQ2}
{
As a next step, we experiment with an alternative, tuned prompt that combines chain-of-thought, few-shot, and role prompting strategies (described in Section~\ref{sec:prompt-engineering}).
We collect and analyze these results to address \textbf{RQ2}.
}

\begin{quote}
\begin{mdframed}[roundcorner=10pt]
    {\bf RQ2:} { Can MLLMs achieve greater-than-human accuracy in identifying lower-stress node-link diagrams?}
\end{mdframed}
\end{quote}

\paragraph*{Results}

\begin{figure*}[ht]
    \includegraphics[width=0.32\linewidth]{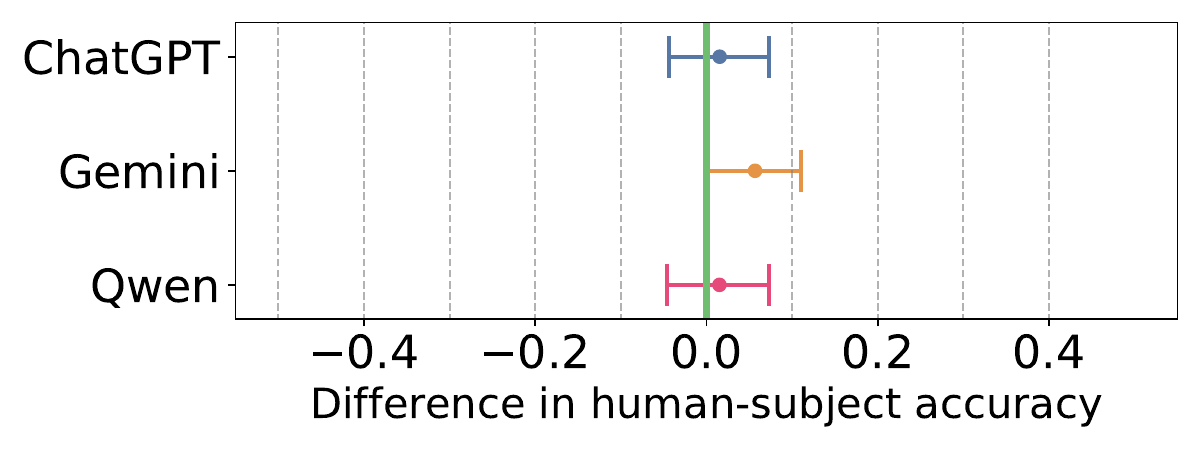}
    \includegraphics[width=0.32\linewidth]{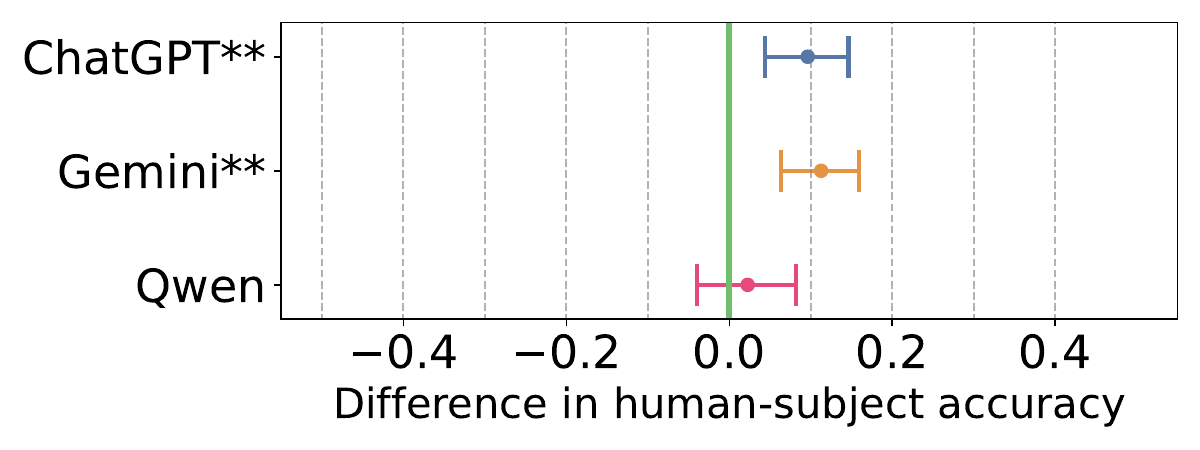}
    \includegraphics[width=0.32\linewidth]{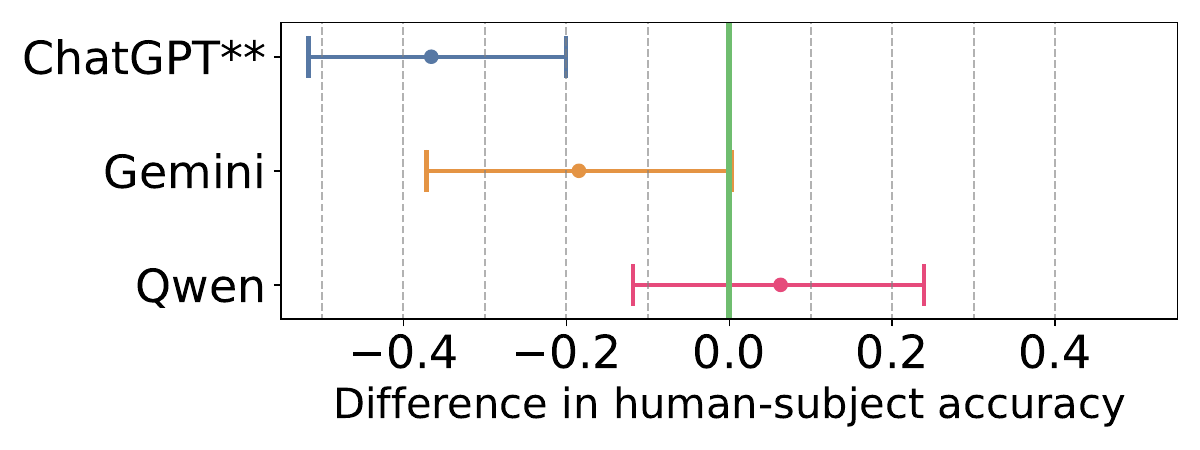}

    \includegraphics[width=0.32\linewidth]{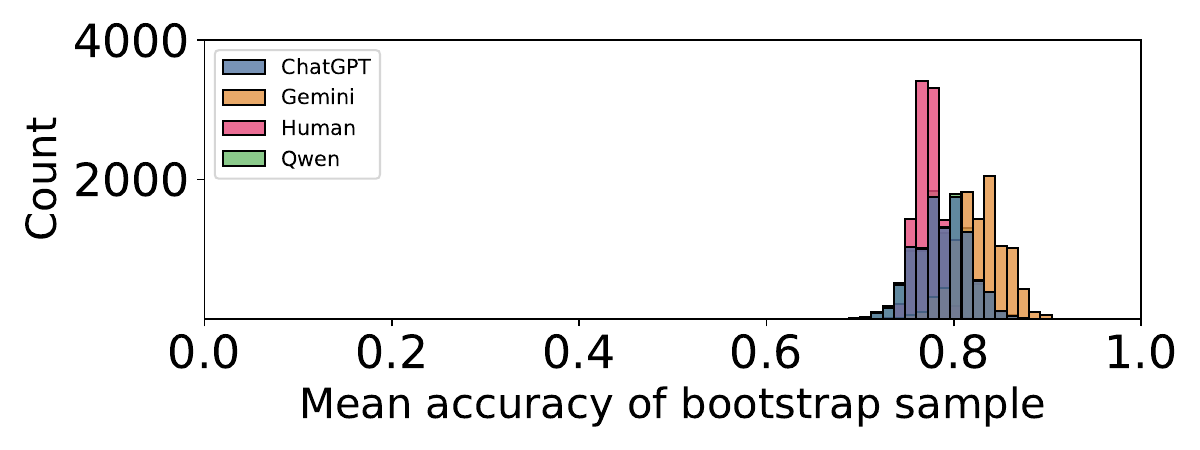}
    \includegraphics[width=0.32\linewidth]{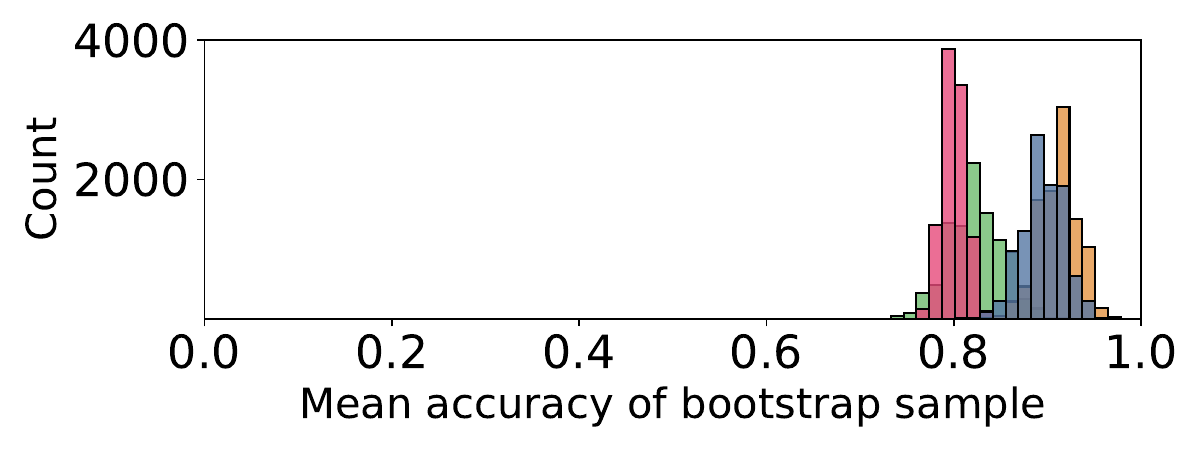}
    \includegraphics[width=0.32\linewidth]{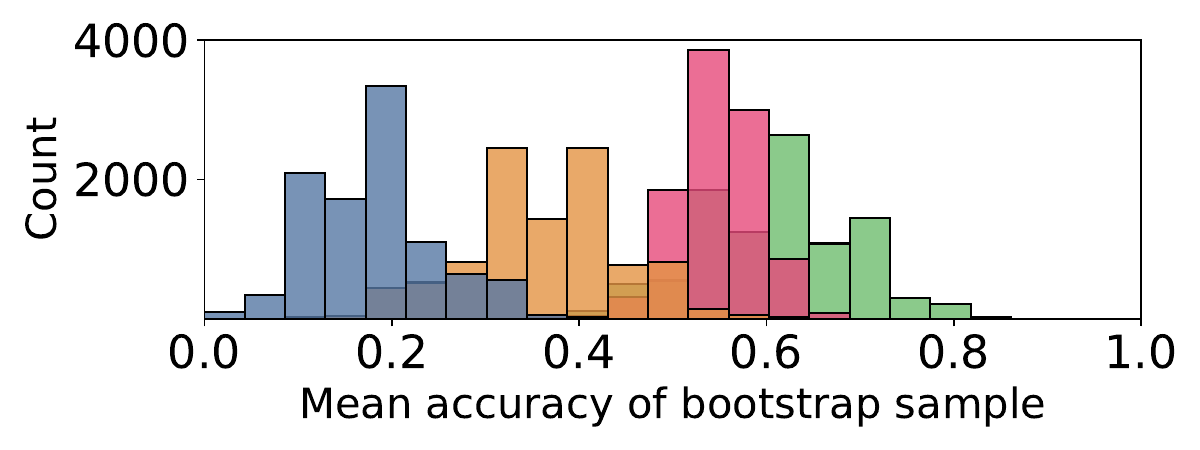}

    \parbox[c]{0.32\linewidth}{\centering (a) \tuned All Levels}
    \parbox[c]{0.32\linewidth}{\centering (b) \tuned Positive Difference}
    \parbox[c]{0.32\linewidth}{\centering (c) \tuned Zero Difference}    

    \caption{(Top) Confidence intervals for the difference between \tuned MLLM and \expert \humansubjects accuracy in the style of Figure~\ref{fig:rq1_DoM}. (a) includes all responses across stress difference levels, (b) includes only positive difference levels, and (c) only includes trials with a difference of zero.
    We see that while the \tuned MLLM setting matches overall accuracy (a) of \expert \humansubjects, MLLMs are statistically more accurate when equal stress pairs are excluded%
    ~(b). The opposite is true when only equal stress pairs are considered (c).
    { A * indicates $p$-value $< 0.05$, ** indicates $p$-value $<0.01$.}
    }
    \label{fig:rq2}
    
\end{figure*}

{
We refer to this setting of model instructions as \tuned. We compare the results of the \tuned prompts to the best performing \humansubject ~group, the \expert participants. We use the same analysis methods as discussed for \textbf{RQ1}. The results are shown in Fig.~\ref{fig:rq2}.
By looking at the overall accuracy, it seems that each of the models performs similarly to \expert \humanparticipants. While they achieve slightly higher accuracy overall, the differences are not statistically significant. However, it is worth mentioning that the confidence interval for \gemini ~only just contains zero; see Fig.~\ref{fig:rq2}(a).

However, there is a clear outlier in 
 the mean accuracy per stress level difference shown in Table~\ref{tab:merged}, 
 when stress difference is zero (when two diagrams have the same stress).
For the positive stress difference case, we observe significant results for both \chatgpt and \gemini, with about 10\% higher accuracy than \expert \humanparticipants. In the zero-stress difference case, we observe a significant difference between \expert \humanparticipants and \chatgpt, but not for  \qwen.

\paragraph*{Discussion}
{ Both \chatgpt ~and \gemini ~models become more accurate than \humanparticipants ~in identifying lower-stress diagrams when there is a non-negligible difference in their stress values.} However, the MLLMs seem reluctant to respond with ``the stress is the same'' (third) option.
Meanwhile, \qwen will often choose this third option and achieve near-human accuracy in the zero-difference setting. One possible explanation is that the \qwen model is smaller but broader in its internal knowledge. This might lead the model not to ``overthink'' the word choice of ``the stress is the same,'' while the other models may reason that this option is unlikely. 
}

Mooney et al.~\cite{DBLP:conf/gd/MooneyPWK024} conjecture in their study that this third option was selected by participants when they were not confident in their answers, which could partially explain the trend of lowering accuracy from stress differences of $0$ to $0.05$ in \human. Interestingly, we do not observe this trend as strongly for the MLLMs. This may be because the MLLMs are ``confident'' in their choice.
It is worth noting that two diagrams in the zero-difference category do not have precisely the same stress value but are within one one-thousandth of each other. There is no evidence that the MLLMs are selecting the true lower-stress diagram (an analysis of this hypothesis using a binomial test is in the supplemental material).

We again remark that it may not be wise to replace traditional human participants for evaluation studies. 
In this case, it is clear that an MLLM can outperform \humansubject ~\experts at this high-level task (with statistically significant differences) and so basing an evaluation on an MLLM would not give a good indication of the usability of a tool, technique, or visualization for a human demographic.

\subsection{RQ3}\label{sec:RQ3}
We start our analysis of RQ3 by establishing whether there is a correlation between stress and three other diagram properties. 
Specifically, we consider node uniformity, edge length deviation, and crossings as visual proxies for stress.

\begin{quote}
\begin{mdframed}[roundcorner=10pt]
    {\bf RQ3:} Is there a relationship between the properties of diagrams and the given answers by the MLLMs?
\end{mdframed}
\end{quote}

\paragraph*{Results}
 Node uniformity~\cite{TaylorR05} is calculated by assuming a $\sqrt{|V|} \times \sqrt{|V|}$ grid over the diagram and computing the spread of nodes. 
A value of 1 indicates that all nodes are in a single cell, while 0 indicates a uniform distribution.
Edge length deviation~\cite{AhmedLDKL22} is the standard deviation from the average edge length. 
Edge crossings~\cite{PurchaseCJ95} is a number between 0 and 1, corresponding to the number of crossings in the diagram divided by the maximum possible (when every edge crosses every other edge).
These three properties
are visual proxies 
frequently mentioned by the participants in the \humansubject ~stress-perception experiment~\cite{DBLP:conf/gd/MooneyPWK024}.

We first perform a Spearman correlation test on all networks in our dataset, showing a significant correlation ($p < 0.01$) between stress and all three proxies. 
Both edge lengths ($\rho = 0.76$) and node uniformity ($\rho = 0.70$) strongly correlate with stress, while crossings ($\rho=0.58$) have a high-moderate correlation
\begin{table*}[t]
\centering
\sffamily
\tiny
\renewcommand{\arraystretch}{1.25}
\caption{Reasoning themes reported by MLLMs in the \textbf{Expert} setting.}
\label{tab:expert-themes}
\begin{tabular}{
>{\columncolor[HTML]{FFFFFF}\centering}m{0.2\linewidth}
>{\columncolor{color1!30}}c 
>{\columncolor{color2!30}}c 
>{\columncolor{color4!30}}c 
>{\columncolor[HTML]{FFFFFF}\centering}m{0.2\linewidth} 
>{\columncolor{color1!30}}c 
>{\columncolor{color2!30}}c
>{\columncolor{color4!30}}c }
\cellcolor[HTML]{343434}{\color[HTML]{EFEFEF} }                                                   & \multicolumn{3}{c}{\cellcolor[HTML]{343434}{\color[HTML]{EFEFEF} \textbf{Proportion Mentioned}}}                                 & \cellcolor[HTML]{343434}{\color[HTML]{EFEFEF} }                                                    & \multicolumn{3}{c}{\cellcolor[HTML]{343434}{\color[HTML]{EFEFEF} \textbf{Proportion Mentioned}}}                                 \\
\multirow{-2}{.275\linewidth}{\cellcolor[HTML]{343434}{\color[HTML]{EFEFEF} \centering  \textbf{Criteria For\\\centering Low Stress}}} & \cellcolor{color1}{\color[HTML]{EFEFEF} \textbf{GPT-4o}} & \cellcolor{color2}{\color[HTML]{EFEFEF} \textbf{Gemini-2.5}} & \cellcolor{color4}{\color[HTML]{EFEFEF} \textbf{Qwen2.5}} & \multirow{-2}{.275\linewidth}{\cellcolor[HTML]{343434}{\color[HTML]{EFEFEF} \centering  \textbf{Criteria For\\\centering High Stress}}} & \cellcolor{color1}{\color[HTML]{EFEFEF} \textbf{GPT-4o}} & \cellcolor{color2}{\color[HTML]{EFEFEF} \textbf{Gemini-2.5}} & \cellcolor{color4}{\color[HTML]{EFEFEF} \textbf{Qwen2.5}} \\
\multicolumn{8}{c}{\cellcolor[HTML]{DDDDDD}\textbf{Related to Node Distribution}}                                                                                                                                                                                                                                                                                                                                                                                            \\
\cellcolor[HTML]{FFFFFF}                                                                          & \cellcolor{color1!30}                                        & \cellcolor{color2!30}        
& \cellcolor{color4!30}   & Even/Balanced Distribution                                                                         & 0.005                                                           & 0.019    & 0.005    
\\
\multirow{-2}{\linewidth}{\centering \cellcolor[HTML]{FFFFFF}Even/Balanced\\\centering Distribution}                             & \multirow{-2}{*}{\cellcolor{color1!30}0.630}                 & \multirow{-2}{*}{\cellcolor{color2!30}0.574}         
& \multirow{-2}{*}{\cellcolor{color4!30}0.588} 
& Uneven/Unbalanced Dist.                                                           & 0.060                                                           & 0.037      & 0.0.23
\\
Spreadout Layout                                                                                  & 0.218                                                           & 0.056          & 0.505                                                & Stretched/Spreadout Layout                                                                         & ---                                                             & 0.069   & 0.009                                                       \\
Compact Layout                                                                                    & 0.046                                                           & 0.009 & ---                                                          & Compact Layout                                                                                     & 0.009                                                           & ---                              &     0.060                        \\Dense Areas/Clusters/Center                                                                              & 0.042                                                           & 0.134         & 0.056                                                 & Dense Areas/Clusters/Center                                                                               & 0.532                                                           & 0.495 & 0.806                                                          \\
Few Dense Areas/Clusters                                                                          & 0.116                                                           & 0.074      & 0.019                                                    & Vertices Outside Clusters                                                                          & 0.009                                                           & 0.019   & 0.097                                                       \\

Organized Layout                                                                                  & 0.139                                                           & 0.079   & 0.181                                                       & Chaotic/Tangled Layout                                                                                     & 0.037                                                           & 0.056       & 0.259                                                   \\
\multicolumn{8}{c}{\cellcolor[HTML]{DDDDDD}\textbf{Related to Edge Length Deviation}}                                                                                                                                                                                                                                                                                                                                                                                        \\
Consistent Edge Lengths                                                                 & 0.009                                                           & 0.130             & ---                                             & Inconsistent Edge Lengths                                                                & 0.005                                                          & 0.028 & ---                                                          \\
Few Long Edges                                                                                    & 0.028                                                           & 0.005 & ---                                                           &                                                               &                                                           &           &                                                     \\
Many Long Edges                                                                                   & ---                                                             & 0.009    & ---                                                       & \multirow{-2}{*}{Many Long Edges}                                                                                 & \multirow{-2}{*}{0.032}                                                          & \multirow{-2}{*}{0.102}          & \multirow{-2}{*}{0.056}                                                  \\
Many Short Edges                                                                                  & ---                                                             & 0.106      & 0.056                                                    & Many Short Edges                                                                                   & ---                                                             & 0.005          & ---                                                \\
\multicolumn{8}{c}{\cellcolor[HTML]{DDDDDD}\textbf{Related to Crossings}}                                                                                                                                                                                                                                                                                                                                                                                                    \\
Few Crossings                                                                                     & 0.968                                                           & 0.421   & 0.537                                                       &                                                                              &   & &                                             \\
Equal Distribution of Crossings                                                                   & ---                                                             & 0.005       & ---                                                   & \multirow{-2}{*}{Many Crossings}                                                                                     & \multirow{-2}{*}{0.829}                                                           & \multirow{-2}{*}{0.282}               &  \multirow{-2}{*}{0.819}                                                               \\
\multicolumn{8}{c}{\cellcolor[HTML]{DDDDDD}\textbf{Other Readability Criteria}}                                                                                                                                                                                                                                                                                                                                                                                              \\
Clarity/Readability                                                                               & 0.069                                                           & 0.079   & 0.334                                                       & Bad Readability                                                                                    & 0.014                                                           & 0.079             & 0.273                                             \\
Low Clutter                                                                                       & 0.028                                                           & ---       & 0.273                                                     & Visual Clutter                                                                                     & 0.315                                                           & 0.148             & 0.472                                             \\
Low Visual Complexity                                                                             & 0.051                                                           & ---    & 0.074                                                        &                                                                              &                                                           &     &                                                    \\
"Straightforward" Layout                                                                          & 0.009                                                           & ---                  & ---                                          &                                                                                                    &                                      &            &                         \\

Low "Density"                                                                                     & 0.042                                                           & ---      & 0.009                                                    &  \multirow{-3}{*}{High Visual Complexity}                                                                             & \multirow{-3}{*}{0.079}                                                           & \multirow{-3}{*}{0.019}           & \multirow{-3}{*}{0.560}                                                    \\
Similar Angles                                                                                    & ---                                                             & 0.005    & ---                                                      & Varying Angles                                                                                     & ---                                                             & 0.005            & ---                                              \\

\multicolumn{8}{c}{\cellcolor[HTML]{DDDDDD}\textbf{Regular Arrangement}}                                                                                                                                                                                                                                                                                                                                                                                                     \\
Linear                                                                                            & 0.005                                                           & 0.009                                                          &   0.028                                                                                                &     Linear &--- &--- & 0.009
\\
Circular/Radial                                                                                   & ---                                                             & 0.014                                                          &     0.032 & Radial & --- & --- & 0.023                               \\
Tree-Like                                                                                         & ---                                                             & 0.009                                                          &     ---                                                                                               & Star-Like                                      &  --- & --- & 0.019                                        \\
\multicolumn{8}{c}{\cellcolor[HTML]{DDDDDD}\textbf{Technical Problems}}                                                                                                                                                                                                                                                                                                                                                                                                      \\
Decision does not reflect reason                                                             & ---                                                             & 0.014                                                          & 0.080 & {\cellcolor[HTML]{FFFFFF}}                                                       & \multicolumn{1}{l}{\cellcolor[HTML]{FFFFFF}}                    & \multicolumn{1}{l}{\cellcolor[HTML]{FFFFFF}}                   & \multicolumn{1}{l}{\cellcolor[HTML]{FFFFFF}}                  
\end{tabular}
\end{table*}

We report results about the potential predictive power in \cref{tab:prediction}. 
The first row of the table shows the accuracy when either a single proxy or a combination of proxies is used to determine which of the two input images has lower stress. By picking the image with a better value in one of the three proxies, one could correctly guess the image with lower stress in more than 84\% of the image pairs. Two or more proxies can be used to predict the lower-stress image, with almost perfect accuracy. As with stress itself, the MLLMs cannot directly compute the values for the three proxies (and thus achieve near-perfect accuracy). However, by systematically assessing multiple aspects, MLLMs are more likely to make the correct decision. 

\begin{table*}[t]
\centering
\sffamily
\small
\setlength{\tabcolsep}{2.8pt}
\caption{{The first row shows the probabilistic predictive power when relying on individual or combinations of properties, such as node uniformity (NU), edge length deviation (ELD), and crossings (XR). The remaining rows show the conditional probability of the
\textbf{trained }model's answer being correct given that the chosen drawing indeed had a better score for that proxy (a better score on all proxies in the case of two or more).
}}
\begin{tabular}{rccccccc}
\rowcolor[HTML]{343434}
\color[HTML]{EFEFEF}Property & \color[HTML]{EFEFEF}NU    &\color[HTML]{EFEFEF} ELD    & \color[HTML]{EFEFEF}XR    & \color[HTML]{EFEFEF}NU\&ELD & \color[HTML]{EFEFEF}NU\&XR & \color[HTML]{EFEFEF}ELD\&XR & \color[HTML]{EFEFEF}NU\&ELD\&XR \\ 
\rowcolor[HTML]{EFEFEF}
\cellcolor[HTML]{343434}{\color[HTML]{EFEFEF}Accuracy} & 0.843 & 0.893 & 0.876 & 1.000 & 0.986 & 0.966 & 1.000

\\ \hline
\rowcolor[HTML]{EFEFEF}
\cellcolor{color1}{\color[HTML]{EFEFEF} \textbf{GPT-4o}}  &
0.873 & 0.932 & 0.914 & 0.959 & 0.932 & 0.943 & 0.955
\\
\cellcolor{color2}{\color[HTML]{EFEFEF} \textbf{Gemini-2.5}}  &
0.931 & 0.942 & 0.952 & 0.973 & 0.973 & 0.966 & 0.970
\\
\cellcolor{color4!60}{\color[HTML]{EFEFEF}\textbf{Qwen2.5}}  &
0.783 & 0.767 & 0.810 & 0.813 & 0.840 & 0.828 & 0.866

\end{tabular}
\label{tab:prediction}
\end{table*}

{
{We conducted our analysis by computing the conditional probabilities of the model's answer being correct given that the chosen drawing indeed had a better score for that proxy (a better score on all proxies in the case of two or more). }
The results for the \textbf{Trained} setting are shown in \cref{tab:prediction}. 
Among the models, \gemini consistently demonstrates the highest conditional accuracy, reaching up to $0.973$ on combined proxies, and shows strong alignment with all proxy signals. \chatgpt follows closely, with comparably high performance across most conditions. In contrast, \qwen performs notably lower, especially on individual proxies such as ELD and XR. Importantly, since all proxies perform well above chance (0.5 for binary selection), we can reasonably infer that the proxies encode meaningful visual cues related to layout stress and that all models leverage these signals. 
However, while above-chance performance supports the proxies' predictive value, it does not imply causality, and some proxy combinations may reflect overlapping rather than additive information. 
}

\paragraph*{Discussion}
Qualitative answers from the human-subject experiment~\cite{DBLP:conf/gd/MooneyPWK024} indicate that humans rely on visual proxies. 
We conjecture that evaluating even a fraction of all possible node pairs would require excessive cognitive load, whereas alternative visual proxies are more manageable to perceive and process.
For example, evaluating stress requires correctly identifying the shortest paths for every pair of nodes, whereas node distribution involves considering a linear number of nodes.

{
The statistical analysis does not provide evidence that MLLMs compute stress exactly. Their reasoning responses describe visual proxies similar to those used by human subjects.
}
Even though we observed this effect across all prompts, it is especially evident with the \expert prompt, which provides the least information, forcing the model to rely on its own context.
A single proxy alone lacks the predictive power to explain our experimental results, especially in networks with greater stress differences, where the models achieve perfect or near-perfect accuracy; see \cref{tab:prediction}. 
Still, the fact that all three visual proxies are (strongly) correlated with stress requires us to be careful when drawing conclusions. 
Without access to the technical details of the models, we cannot rule out that they consider other means besides the visual proxies we analyzed.

\subsection{RQ4}\label{sec:RQ4}

\begin{table*}[]
\caption{Common reasoning patterns mentioned by each MLLM and \humansubjects across different training conditions.}
\sffamily
\small
\renewcommand{\arraystretch}{1.25}
\centering
\begin{tabularx}{0.9958\textwidth}{>{\centering\arraybackslash}m{.1\textwidth}>{\arraybackslash}m{.195\textwidth}>{\arraybackslash}m{.195\textwidth}>{\arraybackslash}m{.195\textwidth}>{\arraybackslash}m{.195\textwidth}}
\rowcolor[HTML]{343434} 
{\color[HTML]{EFEFEF} \textbf{Model}}                                               & \multicolumn{1}{>{\centering\arraybackslash}m{0.195\textwidth}}{\color[HTML]{EFEFEF} \textbf{Untrained}}                                         & \multicolumn{1}{>{\centering\arraybackslash}m{0.195\textwidth}}{\color[HTML]{EFEFEF}\textbf{Trained}}                                                                                                    & \multicolumn{1}{>{\centering\arraybackslash}m{0.195\textwidth}}{\color[HTML]{EFEFEF} \textbf{Expert}}                                                                              & \multicolumn{1}{>{\centering\arraybackslash}m{0.195\textwidth}}{\color[HTML]{EFEFEF} \textbf{Tuned}}                                       \\
\rowcolor[HTML]{EEEEEE}
\cellcolor{color1!60}{\textbf{GPT-4o}}                     & \multicolumn{2}{>{\centering\arraybackslash}m{0.4134\textwidth}}{shortest paths, distance nodes, evenly distributed, proportional}                                                                                                                                           & overlapping edges, visual stress, (fewer) edge crossings, evenly distributed                                        & edge lengths, node distribution, edge crossings, uniformity of edge lengths \\ 
{\cellcolor{color2!60}{ \textbf{Gemini-2.5}}} & \multicolumn{2}{>{\centering\arraybackslash}m{0.4134\textwidth}}{path lengths, visual distance, long edges, distances between nodes, balanced}                                                                                                                              & evenly distributed, visual stress, nodes evenly distributed                                                         & visual stress, edge lengths, uniform node distribution, edge crossings      \\

\rowcolor[HTML]{EEEEEE}{\cellcolor{color4!60}{ \textbf{Qwen2.5}}} 
& path lengths, distance nodes, lower/higher stress, connecting paths, shortest path
& path lengths, distance nodes, lower/higher stress, connecting paths, distribution nodes
& edge crossings, visual complexity, nodes edges, similar level, 
& visual stress, established criteria, examining images \\

\cellcolor{color3!60}{\textbf{Human Subjects}}              & distance between nodes and node distribution, ‘chaotic’, edge lengths, open space & distance between nodes and node distribution, edge lengths, messiness, symmetry, edge crossings, edge closeness, angles, clusters & line length, distances between nodes, edge crossings, value judgments, i.e., \enquote{looks right} & \multicolumn{1}{>{\centering\arraybackslash}m{0.195\textwidth}}{---} 
\end{tabularx}
\label{tab:reason_patterns}
\end{table*}

We analyze the reasoning responses of the MLLMs for \textbf{RQ4}.

\begin{quote}
\begin{mdframed}[roundcorner=10pt]
    {\bf RQ4:} Can we obtain any insight on the decision-making process of the MLLMs?
\end{mdframed}
\end{quote}

\paragraph*{Results}

Every response from the MLLMs is accompanied by a brief explanation in plain text.
This allows us to analyze the decision-making process by looking for common words and similar patterns in the reasoning. { This assumes the reported reasoning accurately reflects the MLLM’s internal decision-making process, which in practice is unknowable. We later examine the predictive value of these reported explanations.}

\begin{figure}
    \centering
    \includegraphics[width=0.95\linewidth]{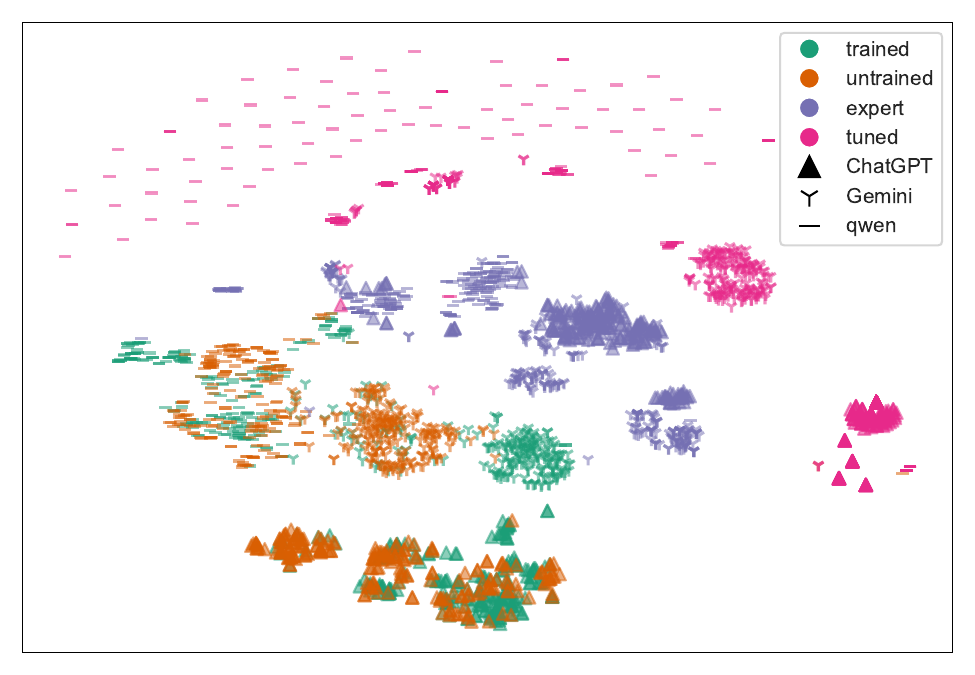}
    \caption{t-SNE projection of the sentence embedding from MLLM responses. We see several well-defined clusters, mostly divided by experimental setting and model. 
    {The answers given by \qwen in the tuned setting are further spread out.}
    }
    \label{fig:overview-rq3}
\end{figure}

We first check for salient differences in MLLM responses between the models and between the three experimental settings. We apply a popular and freely available sentence transformer (sentBERT~\cite{DBLP:conf/emnlp/ReimersG19})
to each of the reasoning responses to obtain a high-dimensional vector representation of each response; see a t-SNE~\cite{JMLR:v9:vandermaaten08a} projection of this data in Fig.~\ref{fig:overview-rq3}. The data is clustered around both of these classes, forming separable clusters for each setting (\trained, \untrained, \expert, \tuned) and model (\chatgpt, \gemini, \qwen). The notable exception is for the \trained and \untrained settings, where there is significant overlap. %
This motivates a thorough analysis of the responses.
We extract themes from the MLLM responses by counting the frequency of phrases and word combinations. { Specifically, we count the occurrence of $n-$grams (length 2--8) in the responses of each model-setting pair, and analyze the most frequently occurring phrases.}
Common reasoning patterns and frequently occurring themes for each setting 
and participant are summarized in \cref{tab:reason_patterns}.
The data of the \humanparticipants is obtained from \cite{DBLP:conf/gd/MooneyPWK024}.
We note that the \humanparticipants were asked about their decision-making process 
after the study, while the MLLMs were asked to provide an explanation after each pair of networks.

It is interesting to observe that the MLLMs in the \untrained and \trained setting 
give very similar explanations for their decisions, 
while they answer differently in the \expert setting.
This could be due to the changed system prompt that tells the MLLM to behave like 
\experts, possibly leading to expert-like language usage. For example, only for the \expert and \tuned setting does \chatgpt ~refer to ``edge crossings'', while in the other settings the phrase  %
``overlapping edges'' is used.

{ In order to analyze the differences between the ``\expert reasoning'' of the models more carefully, we manually labeled the text responses of the MLLMs in the \expert setting according to the reasoning themes used; see \cref{tab:expert-themes}. 
}
In the \expert setting, prompts contain neither the definition of stress nor references to proxies, i.e., we can expect that MLLMs provide reasonings based on their native interpretation of network diagrams. The most noteworthy differences between the MLLMs are the following. First, \chatgpt ~responses almost always contain references to the number of crossings, whereas \gemini ~mentions this only half as often
(this accounts for each reasoning using at least one of the themes ``Few Crossings'', ``Equal Distribution of Crossings'', and ``Many Crossings''). On the other hand, \gemini ~ more often refers to proxies measuring edge-length deviation than \chatgpt does. To this end, it is noteworthy that edge-length deviation correlates strongly with stress, as we discuss in the next section. While both models report clusters of nodes in dense areas as an obstruction for low stress in roughly half of the responses, curiously, both also mention it sometimes as a helpful feature for reducing stress. \gemini ~does that more often in more than 10\% of all its responses. Both models also address several other readability-related properties.
Both models most often mention visual clutter, readability, and visual complexity; the remaining such topics are distinct between the two models. Finally, we also remark that in three cases, \gemini ~provided a reasoning contradicting its chosen answer, e.g., it chose the ``left'' image by choosing Image 1,
but then provided a reasoning stating ``\emph{The drawing on the left has a more centralized structure, with many edges converging on a smaller number of nodes. This creates a sense of congestion and visual stress.}''

{
To further analyze the MLLMs' responses, we align the visual proxies from \cref{sec:RQ3} with the codes derived from the textual answers. We use a subset of the codes in \cref{tab:expert-themes} and combine them into three overarching categories: node uniformity, edge length, and crossings.
Then, we compute the conditional probability of a model being correct when mentioning the category. For example, if a model mentions uniform node distribution in its answer and the drawing numerically corresponds to better node uniformity, then we consider it correct. 
\cref{tab:proxy_text} shows the results for all three models.
We can observe that all models are just slightly better than guessing. This indicates that the actual text responses do not strongly reflect the decision. This is surprising as the models seem to pick up the signal from the visual proxies. 
One explanation could be that the models learn to imitate a good reasoning answer, but do not correctly connect this information to the image. Hence, it infers an explanation that is not grounded in reality. 
}

\paragraph*{Discussion} Both \chatgpt ~and \gemini ~discuss visual proxies correlated with stress in their reasonings --  node distribution, edge length deviation, and number of edge crossings. 
This happens even in the \textbf{Expert} setting where the definition of stress is not provided. 
The two models appear to value aesthetic criteria differently in the \expert setting, where the least amount of additional information is provided via prompts. \chatgpt ~ focuses on the presence of crossings, followed by node distribution. In contrast, \gemini ~mentions node distribution more often than reasons related to the number of crossings and also considers the edge length deviation in 26.9\% of all cases compared to 5.6\% for \chatgpt. 
{Although the models mention visual proxies,
there is no strong  correlation to the actual shown images.
}

Only when provided with a definition (in conditions \untrained and \trained),  both \chatgpt ~and \gemini ~consistently refer to patterns related to the definition of stress (``path lengths''/``shortest paths'', ``distance nodes''/``proportional'', ``distances between nodes'').
In contrast, in the \expert and \tuned setting,  both models often refer to the term ``{visual stress}''. 
Proportionality between distances and lengths of shortest paths is often used as an ending of sentences by most models, e.g., by saying \emph{``Image 1 appears to have a more even distribution of nodes and edges, suggesting that the distances between nodes are more proportional to the shortest paths.''} (\chatgpt ) or \emph{``Image 1 has a more uniform distribution of nodes and edge lengths, suggesting a better correspondence between distance and path length, hence lower stress.''} (\gemini ). Given that the MLLMs are not allowed to execute code 
(verified by API limitations), it is unlikely that they in fact computed stress according to the definition, and this rather indicates that both models are eager to express that they understood the assignment. 

Overall,  we can extract insights on the decision-making process, but careful prompting and analysis are required as both models tend to mention phrases provided in prompts even if they are not used in the actual decision process.

\begin{table}[t]
\centering
\sffamily
\small
\setlength{\tabcolsep}{2.8pt}
\caption{{Conditional probability of the models mentioning a reasoning theme in the \textbf{Expert} setting and the answer correctly aligning with a visual proxy.}}
\begin{tabular}{rccccccc}
\rowcolor[HTML]{343434}
\color[HTML]{EFEFEF}Property & \color[HTML]{EFEFEF}NU    &\color[HTML]{EFEFEF} ELD    & \color[HTML]{EFEFEF}XR    \\

\rowcolor[HTML]{EFEFEF}
\cellcolor{color1}{\color[HTML]{EFEFEF} \textbf{GPT-4o}}  &
0.526 & 0.500 & 0.535\\
\cellcolor{color2}{\color[HTML]{EFEFEF} \textbf{Gemini-2.5}}  &
0.593 & 0.490 & 0.564\\
\cellcolor{color4!60}{\color[HTML]{EFEFEF}\textbf{Qwen2.5}}  &
0.529 & 0.636 & 0.561

\end{tabular}
\label{tab:proxy_text}
\end{table}

\section{Discussion and Limitations}

Our findings suggest that MLLMs perform similarly to \humansubjects ~on a perceptual task involving node-link diagrams, with relatively small but statistically significant deviations based on the amount of context given. This is in spite of recent work~\cite{DBLP:journals/tvcg/BendeckS25,Wang2025} which indicates MLLMs may struggle with complex visualizations, but is in line with the general upward trajectory of MLLM performance on increasingly complex tasks. Additionally, the MLLMs give reasonably sound justifications for their choices and seem to use similar visual proxies to \humanparticipants, but it remains unclear exactly how they arrive at a conclusion.

Our results have practical implications for design, development,~and implementation in future visualization evaluation studies. We recommend care in such experiments,
using attention checks and carefully considering how participants could use MLLMs that may skew study results. In particular, the near-human accuracy of MLLMs suggests that designers must account for the possibility that such models could be participating in human-subjects studies, as pointed out by Agnew et al.~\cite{DBLP:conf/chi/AgnewBCDEPMM24}. Thus, as proposed by H{\"{a}}m{\"{a}}l{\"{a}}inen et al.~\cite{DBLP:conf/chi/HamalainenTK23}, usage of LLMs and MLLMs, should be limited to pilot experiments. 

We emphasize that our experiments do not directly aim to draw~new conclusions on the human perception of network visualizations, { nor do we intend for MLLMs to be deployed to solve perceptual tasks.} However, our results may provide early evidence that MLLMs may indeed be applicable in studies on the aesthetics of (relational) data visualizations, with some caveats. 
For example, the MLLMs consistently underused the option, ``the stress is the same.'' The use of the word \enquote{same} is potentially confounding, as it is slightly misleading; two network diagrams are unlikely to have exactly the same stress values. However, two diagrams are considered to have the \enquote{same} stress if they are within a small threshold of each other. Had an MLLM been used as a pilot test subject in this study, this wording issue might have been caught. {The difference in effect of word choice between human-subjects and MLLMs is an interesting avenue for future work. }
{
Additionally, it seems likely that while MLLMs may perform well on global perceptual tasks such as identifying stress, they may struggle more with local, low-level tasks that require examining individual nodes and edges, e.g., finding a shortest path.
}

{
In our replication study, we intentionally use the same stimuli as the target human-subjects experiment. However, several visualization choices are made in displaying the networks, e.g., color, shape, line thickness, etc. It does not seem obvious that choices that benefit human observers also benefit MLLMs and vice versa. Do best practice encoding standards still hold for MLLMs?
}

{
When prompting the MLLM for a response, we ask for an answer, an explanation, and confidence in this order to match the format of the reference study. Many recent prompting strategies now take a ``reasoning first'' approach, which tends to increase accuracy. We ran a small-scale experiment testing this against the order used in our study. However, we observe slightly higher accuracy in the reporting ``reasoning second'' approach we used. Further details are found in the supplemental material.
}

{
Both \chatgpt ~and \gemini ~are closed-source, making replication of studies difficult as these models are phased out. However, the inclusion of \qwen helps alleviate this concern. 
}
We investigate only one high-level perceptual task. While we investigate this task in great detail, it does not tell us much about the performance of other specific tasks on complex visualizations. We note that evaluating the stress of a node-link diagram involves several low-level tasks (path-following, distance estimation, value derivation, etc.) and high-level perceptual processes (estimating symmetry, distribution of points and lines). 
{
Additionally, our tuned prompt employs a set of established prompting strategies (chain-of-thought, few-shot, role, and comparative prompting).
More sophisticated techniques, such as self-consistency~\cite{WangWSLCNCZ2023} or tree of thoughts~\cite{YaoYZS00N23}, which sample or explore multiple reasoning paths, could further improve accuracy, particularly for the zero-difference stimuli where the models struggle most.
However, these techniques require multiple API calls per stimulus, substantially increasing cost.
We leave such exploration to future work.
}
Finally, although we attempt to replicate the \humansubject ~study of Mooney et al.~\cite{DBLP:conf/gd/MooneyPWK024} as precisely as possible with MLLMs, there are small differences in how the data is given and collected (e.g., every request to the MLLM is independent). 

{
Our study shows that even relatively minor choices in prompting can have a non-trivial effect on results (e.g., differences in accuracy between the untrained and expert settings). Due to the opaque nature of MLLMs, no explanation is available for this difference, but further exploration is a good avenue for future work.
} 
There are many interesting avenues for future work involving MLLMs and node-link diagrams.
To this end, all prompts, code, data, results, and analysis are provided in an open-source OSF repository: \url{https://osf.io/748mx/}.

\section{Conclusions}
We investigate a topical question in visualization: ``Do MLLMs have the perceptual capacity to understand complex visualizations?'' We attempt to answer this question by replicating a recent human-subject study on a high-level perception task for node-link diagrams, using MLLMs. Our findings indicate that MLLMs achieve results comparable to human performance, and 
by tuning the prompt, MLLMs can outperform humans. %
Reasoning responses from MLLMs provide some insight into their underlying decision processes, including evidence that they use visual proxies similar to those used by humans.
However, analysis of the textual answers shows that the provided reasoning does not necessarily correspond to the decision.

\section*{Acknowledgments}{%
	We thank Gavin Mooney for the stimuli generator and the human-subject responses of the reference experiment.
}

\bibliographystyle{abbrv-doi-hyperref-narrow}
\bibliography{shortened_bibliography}

\appendix

\section*{Experimental data}
All stimuli, experimental scripts, and analysis can be found in our online repository: \url{https://osf.io/748mx/}.

\section*{Full Prompts}

We show here the prompts in full given to the MLLMs. The raw markdown files can be found in our open-source repository.

\paragraph{\textbf{Trained}}
The prompts given in the trained setting were as follows:
\begin{quote}
    
You are a person participating in a study.

There will be a short test at the start of the study to ensure that you understand the relevant concepts. If you do not get more than 50\% for this test, you will not be able to proceed to the paid experiment. You will receive training and examples on the relevant concepts in advance of this test.

The aim of the study is to investigate how well people can see the difference between drawings of networks. The results of this experiment can help guide the design of visualisations of networks, for ease of understanding. We focus on how visually balanced (or `stressed') networks appear.
 
You will be asked to look at two networks, side by side. You will be asked to indicate which one has more `stress'. An explanation of `stress' will be given in advance, together with examples.
 
In brief: Stress in a network drawing is defined as tension between the distance between nodes and the length of the path between them.
 
Since we are interested in the immediate perception of the visual properties, we ask that you make your decision as soon as possible; we do not expect you to examine the drawings in great detail.

\end{quote} 

\paragraph*{\textbf{Untrained}}
The prompts given in the untrained setting were:
\begin{quote}
You are a person participating in a study.

There will be a short training section at the start of the survey.

The aim of the study is to investigate how well people can see the difference between drawings of networks. The results of this experiment can help guide the design of visualisations of networks, for ease of understanding. We focus on how visually balanced (or `stressed') networks appear.
 
You will be asked to look at two networks, side by side. You will be asked to indicate which one has more `stress'. An explanation of `stress' will be given in advance, together with examples.
 
In brief: Stress in a network drawing is defined as tension between the distance between nodes and the length of the path between them.
 
Since we are interested in the immediate perception of the visual properties, we ask that you make your decision as soon as possible; we do not expect you to examine the drawings in great detail.

\end{quote}

\paragraph*{\textbf{Expert}}

The prompts given in the expert setting: 

\begin{quote}
You are an expert in graph and network visualization.

You will be asked to look at two networks side by side. You will be asked to indicate which one has more `stress'.

Since we are interested in the immediate perception of stress, we ask that you make your decision as soon as possible; we do not expect you to examine the drawings in great detail. 

\end{quote}

\paragraph*{\textbf{Tuned}}

The prompts given in the tuned setting: 

\begin{lstlisting}[linewidth=\columnwidth,breaklines=true]
    **You are an expert evaluator of graph visualizations, specifically focused on comparing the visual stress levels of two graph images.** You will receive two graph images (labeled Image 1 and Image 2), from which you will visually assess their nodes, edges, spatial arrangement, and any observed edge crossings. Your evaluation will be based on established principles for visually clear and balanced network drawings.

## Understanding of "Visual Stress" for this Task

For the purpose of this evaluation, "visual stress" refers to qualities in the drawing that hinder readability and comprehension due to visual clutter, imbalance, or tension. It is assessed based on visually perceivable layout characteristics, not a formal mathematical computation. Higher visual stress makes the graph harder to interpret quickly and accurately. We will focus on the following aesthetic criteria:

## Evaluation Criteria and Prioritization

Evaluate the images based on the following criteria. When results are mixed, prioritize them in this order:

1.  **Distribution of Nodes (Highest Priority):** A more uniform and balanced distribution of nodes reduces visual stress. Uneven distribution (dense clusters vs. large empty areas) increases stress.
    *   Assess the *evenness* of node spacing across the entire drawing area for both images.
    *   Identify any significant *clustering* (regions with much higher node density than average, e.g., >25%
    *   Compare the overall spatial balance.
    *   **Lower Stress Indicator:** More uniform node spacing, less clustering, fewer large empty areas.

3.  **Uniformity of Edge Lengths (Medium Priority):** More uniform edge lengths contribute to lower visual stress, though this is less critical than crossings or node distribution. Extreme variations can create visual imbalance.
    *   Visually compare the *range* of edge lengths in both images. Are most edges similar in length, or is there high variability?
    *   Estimate the approximate ratio between the longest and shortest edges.
    *   Note the presence of any significant *outlier edges* (much longer or shorter than the average).
    *   **Lower Stress Indicator:** Less variation in edge lengths, smaller ratio between longest and shortest edges.

3.  **Crossings of Edges (Lowest Priority):** Fewer edge crossings significantly reduce visual stress.
    *   Visually estimate and compare the *number* and *density* of edge crossings in both images.
    *   Note if crossings are concentrated in specific areas or distributed throughout.
    *   Estimate the approximate percentage difference in crossings if one image is clearly better.
    *   **Lower Stress Indicator:** Significantly fewer crossings.



## Evaluation Process

1.  **Examine Both Images:** Carefully observe the layout of nodes and edges in Image 1 and Image 2, paying attention to the overall structure and spatial relationships based on the criteria above.
2.  **Compare Criterion by Criterion:**
    *   **Node Distribution:** Determine which image has a more uniform and balanced node distribution.
    *   **Edge Lengths:** Determine which image exhibits more uniform edge lengths.
    *   **Edge Crossings:** Determine which image performs better regarding the number and density of crossings.
3.  **Synthesize Findings & Prioritize:** Weigh the findings according to the prioritization (Distribution > Lengths > Crossings). For example, a significant advantage in node distribution outweighs minor disadvantages in crossings or edge lengths.
4.  **Determine Overall Lower Stress:** Conclude which image exhibits lower overall visual stress based on the weighted comparison.
5.  **"Same Stress" Condition:** Select option (3) *only* if the images are visually very similar across all criteria, or if advantages in one criterion are clearly offset by disadvantages in another of equal or higher priority, resulting in a negligible overall difference (e.g., visually estimated <10'%

\end{lstlisting}

\paragraph*{\textbf{Unsuccesful Attempts}}

{ We tested multiple variants  of the \tuned prompt with mixed success and report our findings here. Firstly, instead of giving the same examples as in the study, we tried to give specific examples where the models performed poorly. For example, we gave more image pairs that show a network with similar stress. While this improves the accuracy in the 'same stress' bucket, it results in overall lower accuracy due to the decrease in all other buckets. }
Hence, we give the models the same examples as in \cref{sec:prompts}. 

We also tried to see if forcing the models to compute the stress from the image input is possible. Our preliminary results with the web interface indicated that this is potentially feasible, but we could not evaluate this variant due to the API's technical limitations. However, extracting the network structure and computing the stress is impossible even with the web interface. 
The models could not accurately extract a network from the image input, even when explaining how a node or edge is visually represented and correcting mistakes. 
We conjecture that this capability will be possible in the future.

{
\section*{Order of Responses}
We ask the MLLM to give its responses in a fixed order: First the discrete choice for the stimuli (Image 1, Image 2, Neither), second an explanation of why, and third a confidence score. This order was very intentional, as it matches the order of questions given to human-subject participants of the reference experiment. However, it is known that MLLMs tend to perform better with a "reasoning-first" strategy, coming to a better solution after being explicitly asked to reason through the problem. 

However, we do not see such performance gains; likely because the models we deploy are so-called ``reasoning models'' which perform implicit reasoning before providing an answer. We show results of an experiment with the \qwen model in Figure~\ref{fig:reverse-forward}. Here we test the human-subjects like setting of reasoning second (``forward'') against the reasoning-first setting (``reverse''). We see that while differences are relatively small, accuracy tends to be slightly higher overall for the forward setting.

\begin{figure}[ht]
    \centering
    \includegraphics[width=\linewidth]{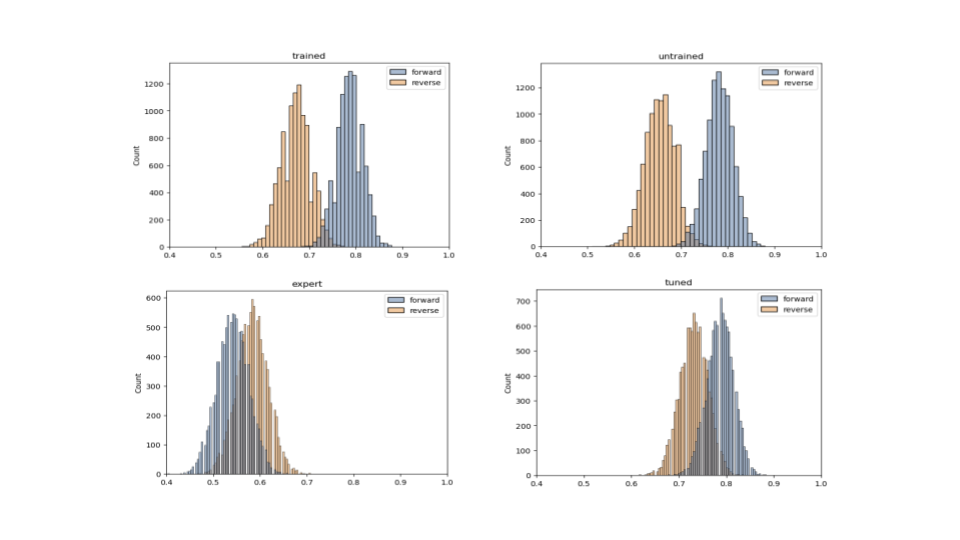}
    \caption{Accuracy of Qwen when prompted to give reasoning after the response (forward) and reasoning first (reverse). Note that while the difference is small, the reverse setting tends to have lower accuracy across settings.}
    \label{fig:reverse-forward}
\end{figure}
}

\section*{Additional data}

\begin{figure*}[t]
    \centering
    \begin{tabular}{c m{0.21\textwidth} m{0.21\textwidth}m{0.21\textwidth}m{0.21\textwidth}}

    & \multicolumn{1}{c}{\textbf{Trained}}
    & \multicolumn{1}{c}{\textbf{Untrained}}
    & \multicolumn{1}{c}{\textbf{Expert}}
    & \multicolumn{1}{c}{\textbf{Tuned}} \\
    
    $\mathbf{n=10}$
    &\includegraphics[width=\linewidth]{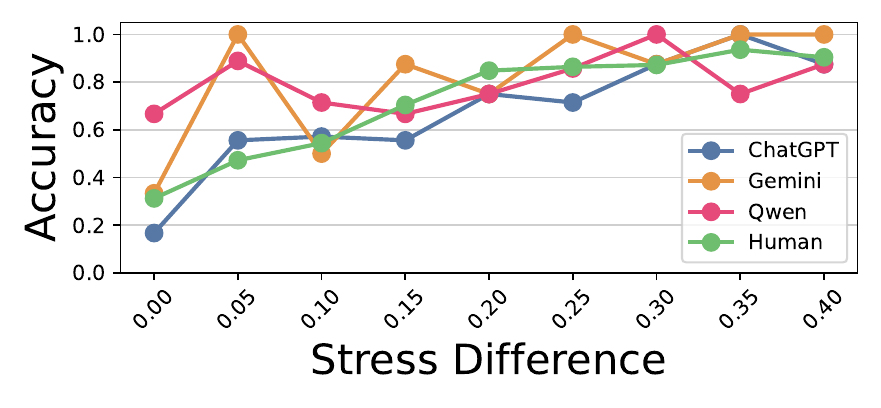}
    &\includegraphics[width=\linewidth]{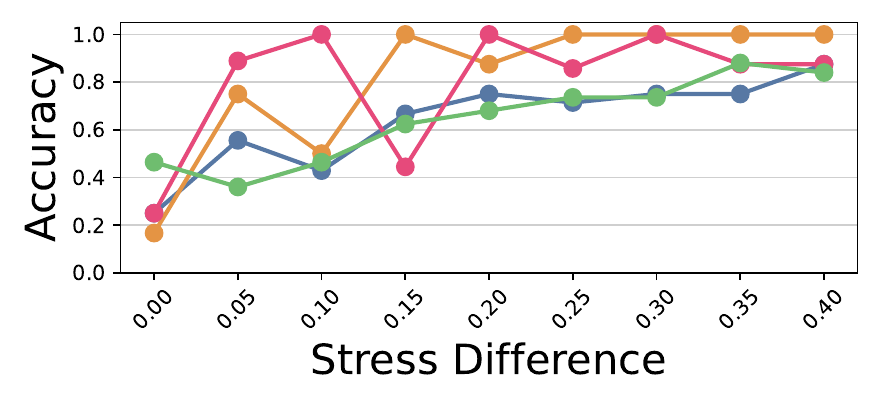}
    &\includegraphics[width=\linewidth]{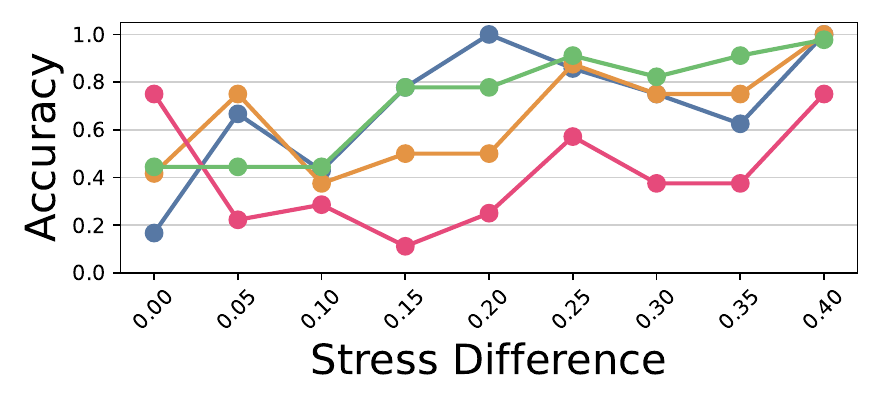}
    &\includegraphics[width=\linewidth]{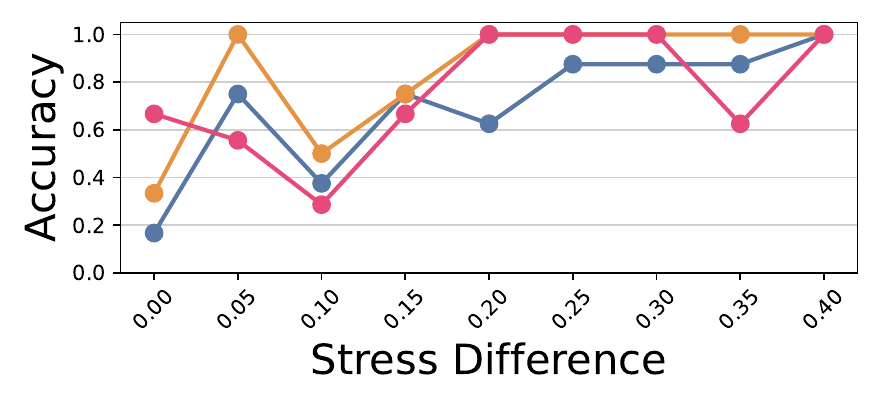}\\

    $\mathbf{n=25}$
    &\includegraphics[width=\linewidth]{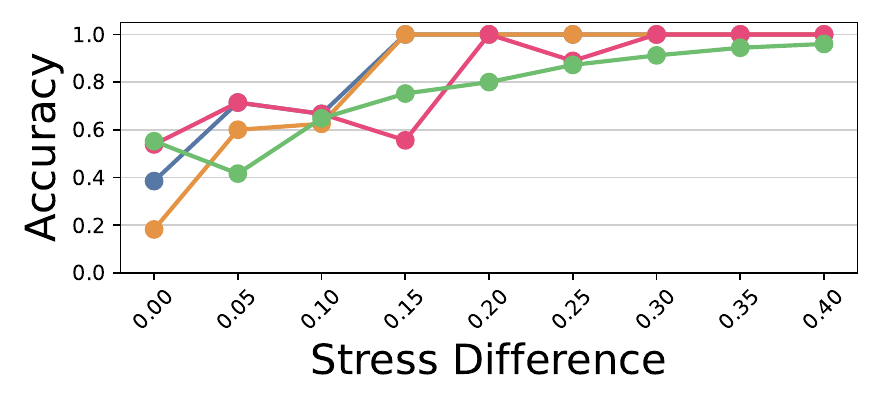}
    &\includegraphics[width=\linewidth]{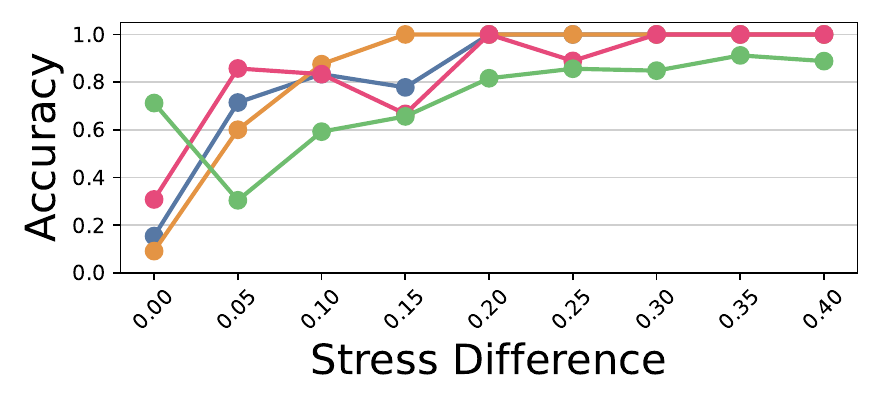}
    &\includegraphics[width=\linewidth]{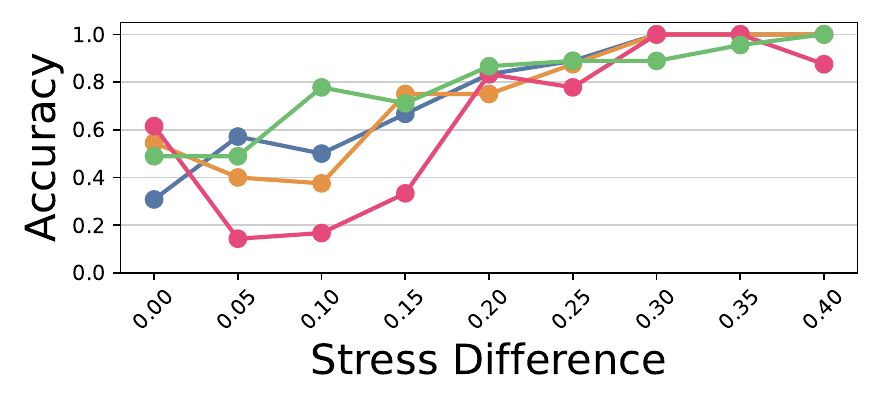}
    &\includegraphics[width=\linewidth]{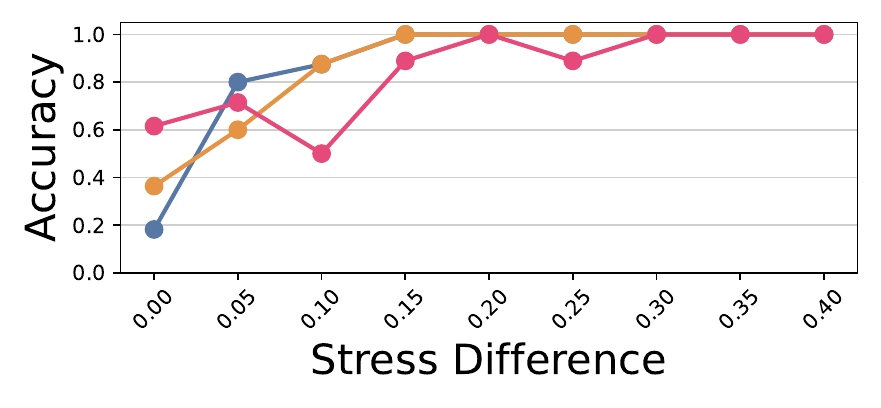}\\

    $\mathbf{n=50}$
    &\includegraphics[width=\linewidth]{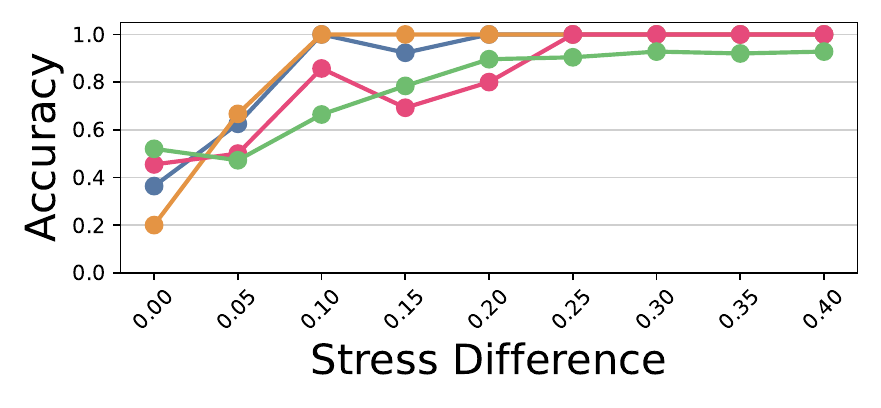} 
    &\includegraphics[width=\linewidth]{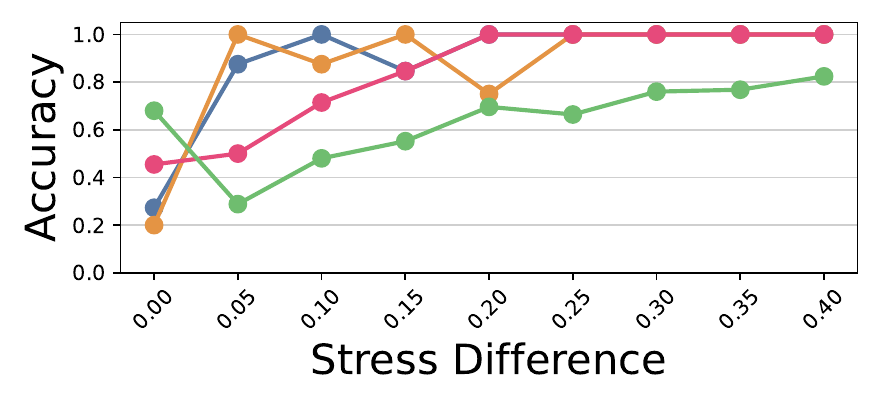}
    &\includegraphics[width=\linewidth]{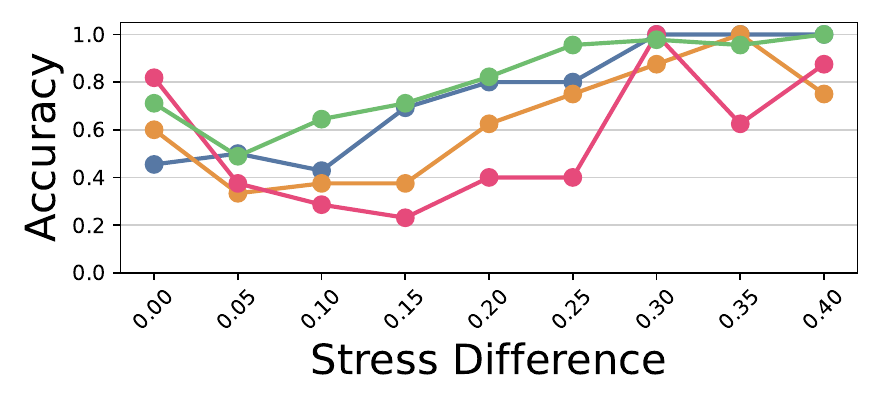}
    &\includegraphics[width=\linewidth]{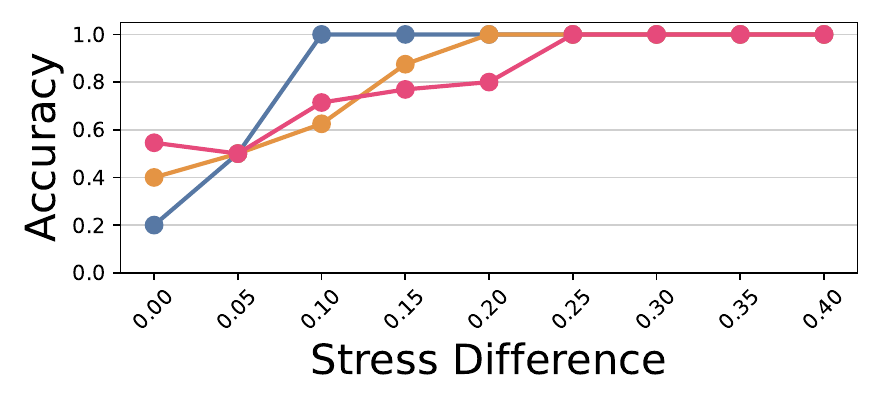}

    \end{tabular}

    \caption{Overall accuracy in the \textbf{Trained}, \textbf{Untrained}, \textbf{Expert} and \textbf{Tuned} 
    setting for \chatgpt, \gemini, \qwen, and \human with respect to stress level difference. Every row represents the size of network seen, and every column a different setting. All trends tend to increase in accuracy as the stress difference gets larger. \chatgpt's are offset by 0.01 as they often overlap with \gemini.}
    \label{fig:linechart2}
\end{figure*}

We include more detailed means for each setting (\textbf{Trained}, \textbf{Untrained}, \textbf{Expert}, and \textbf{Tuned}) and size (10, 25, 50). Each setting and size has data for each model, \chatgpt, \gemini, and \qwen, as well as data for \humanparticipants ~in all but the \textbf{Tuned} setting. This data is visualized in small multiple line charts in Fig.~\ref{fig:linechart2}. All means are readable in Table~\ref{tab:merged:2}.

\begin{figure}[tbh]
    \centering
    \includegraphics[width=0.49\linewidth]{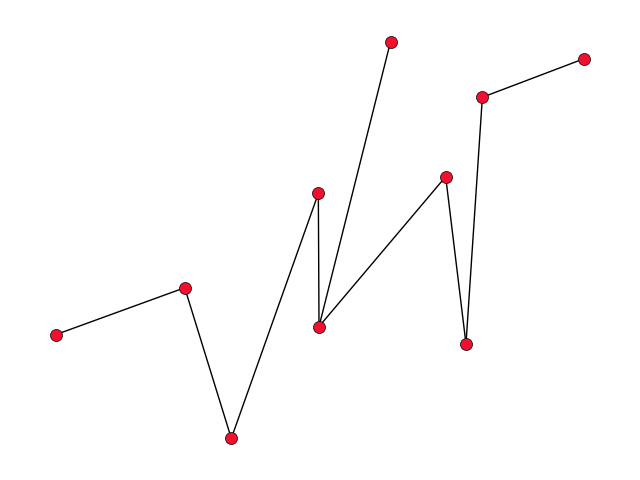}
    \includegraphics[width=0.49\linewidth]{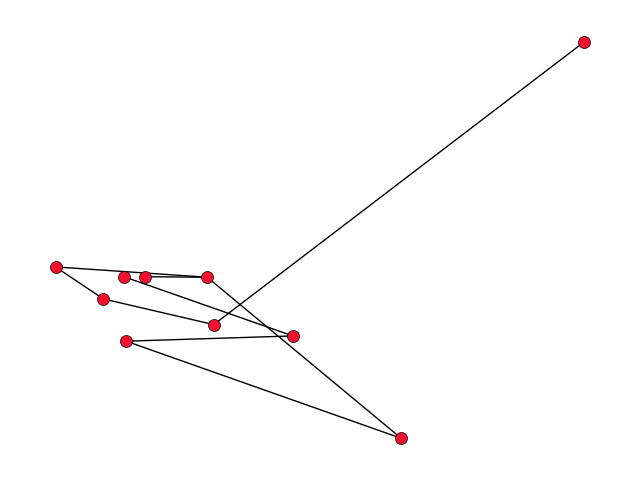}

    \includegraphics[width=0.49\linewidth]{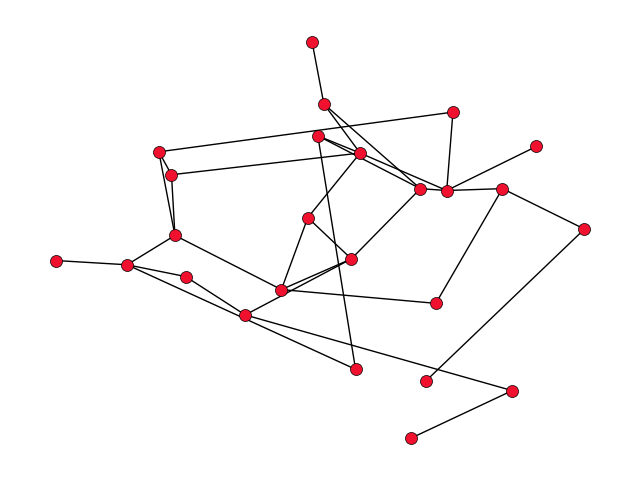}
    \includegraphics[width=0.49\linewidth]{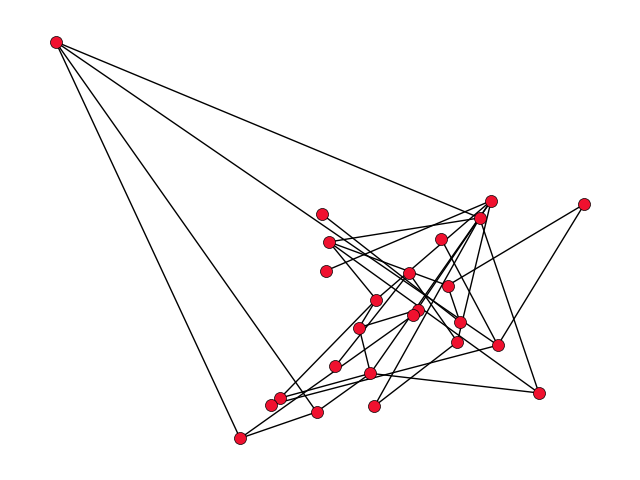}

    \includegraphics[width=0.49\linewidth]{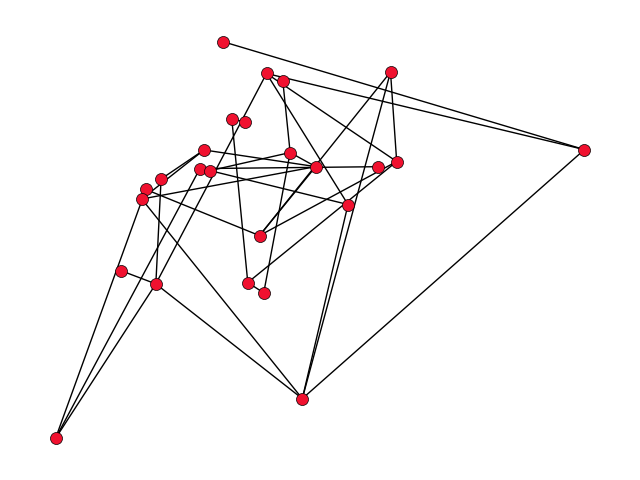}
    \includegraphics[width=0.49\linewidth]{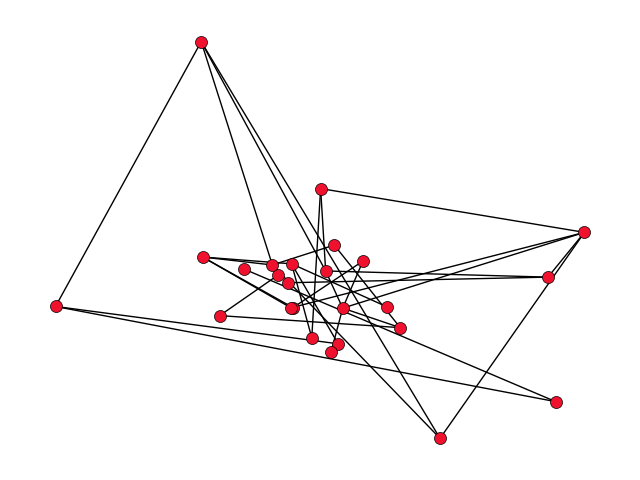}

    \includegraphics[width=0.49\linewidth]{figs/graphs_sm/n25-2A-drawing-0.7.png}
    \includegraphics[width=0.49\linewidth]{figs/graphs_sm/n25-2A-drawing-0.45.png}
    
    \caption{Example stimuli (pairs of network diagrams). In all examples, the lower stress diagram appears on the left.}
    \label{fig:graphs1}
\end{figure}

\begin{figure}[tbh]
    \centering
    \includegraphics[width=0.49\linewidth]{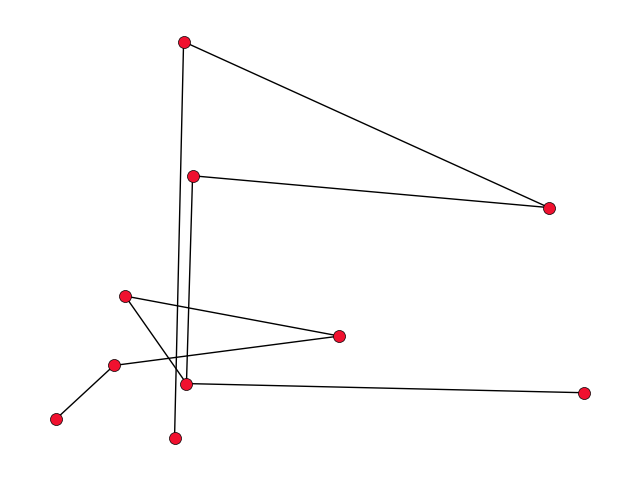}
    \includegraphics[width=0.49\linewidth]{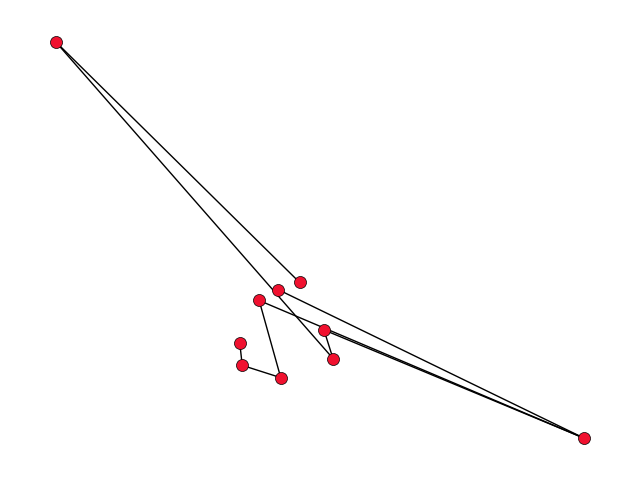}
    
    \includegraphics[width=0.49\linewidth]{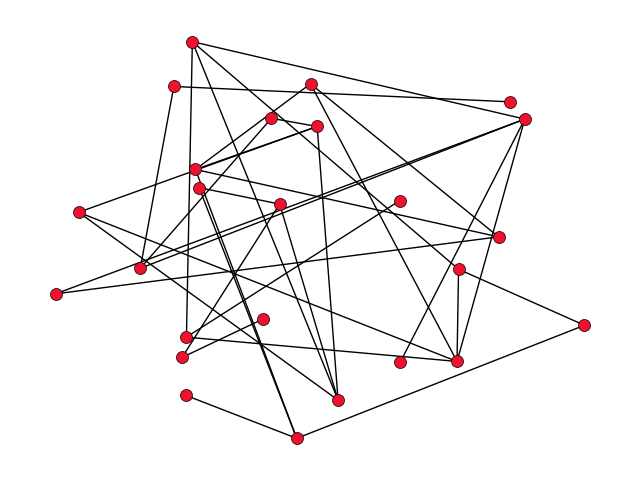}
    \includegraphics[width=0.49\linewidth]{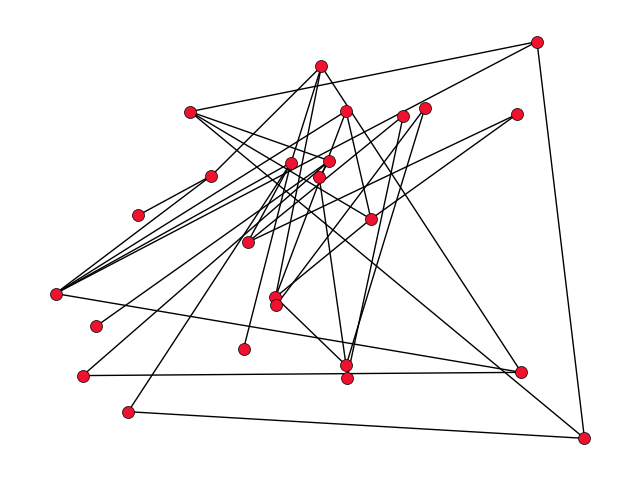}

    \includegraphics[width=0.49\linewidth]{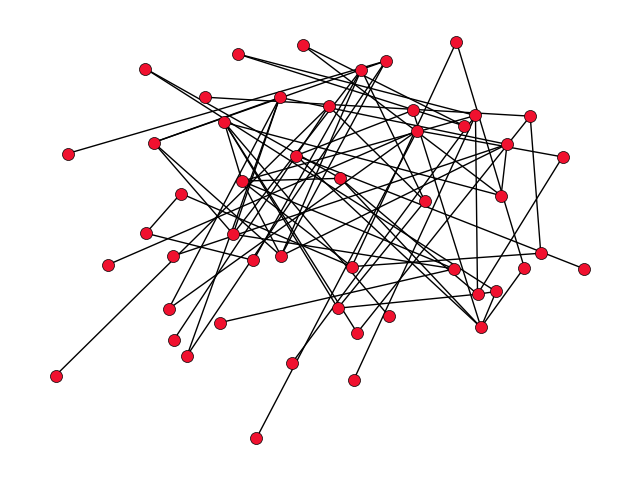}
    \includegraphics[width=0.49\linewidth]{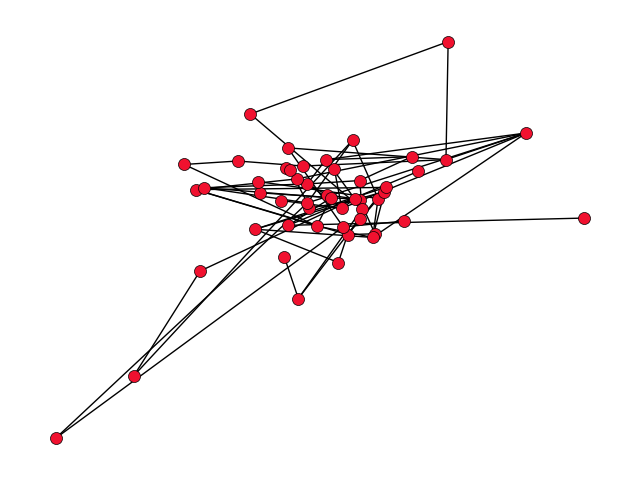}

    \includegraphics[width=0.49\linewidth]{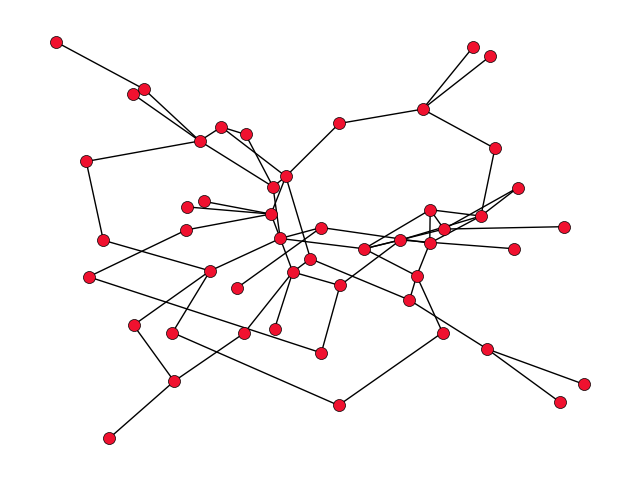}
    \includegraphics[width=0.49\linewidth]{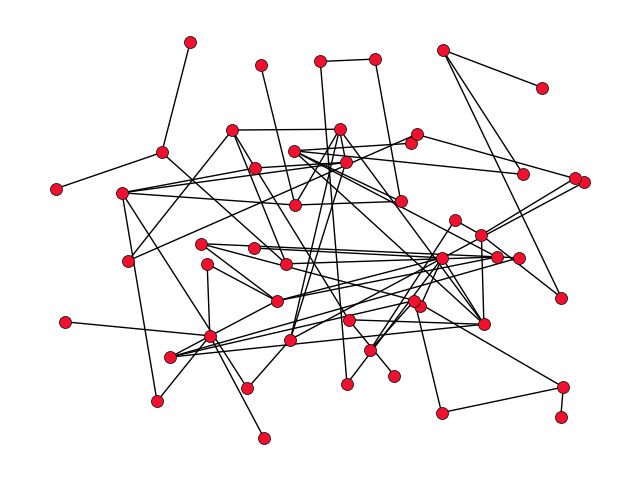}
    
    \caption{Example stimuli (pairs of network diagrams). In all examples, the lower stress diagram appears on the left.}
    \label{fig:graphs2}
\end{figure}

Example stimuli pairs can be found in Fig.~\ref{fig:graphs1} and Fig.~\ref{fig:graphs2}. All stimuli can be found on OSF.

\begin{figure}[tbh]
    \centering
    \includegraphics[width=0.48\linewidth]{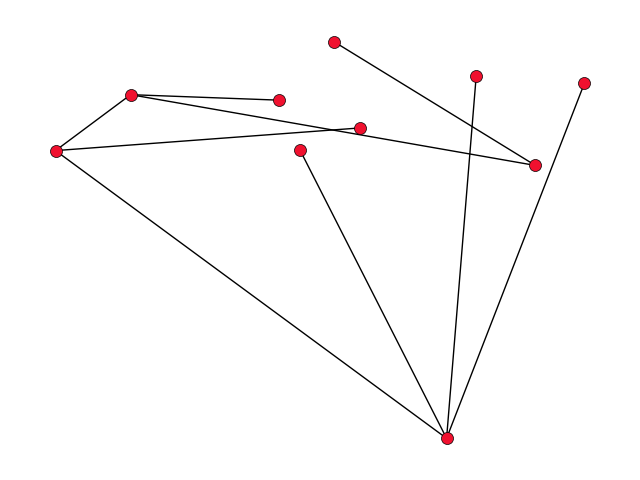}
    \includegraphics[width=0.48\linewidth]{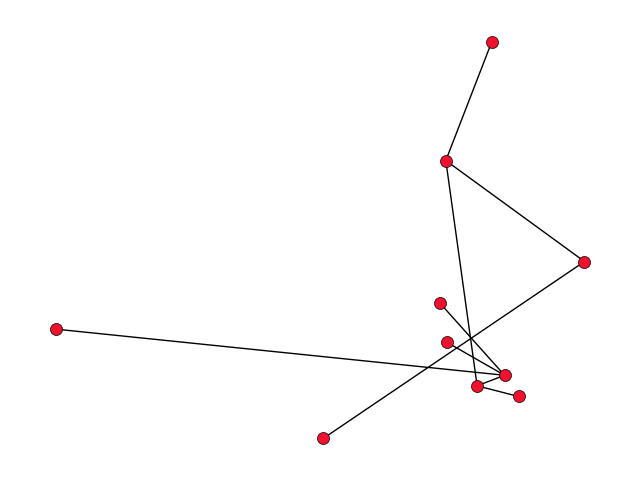}
    \caption{The diagram of the network on the left has a higher stress value of 0.65 compared to 0.5 for the diagram on the right, but the right diagram has worse node uniformity, edge length variation, and more crossings. The outlier node at the bottom of the left diagram is the culprit of stress as other nodes are spatially close, but some paths must go over it.}
    \label{fig:proxy-graphs}
\end{figure}

\clearpage

\begin{table*}[hptb!]
\caption{Aggregated results in the \textbf{Trained}, \textbf{Untrained}, and \textbf{Expert} setting for~\chatgpt,~\gemini,~\qwen, and~\human, and \textbf{Tuned} for the MLLMs.
The rows show the absolute stress level difference between the two network diagrams shown to the MLLM and participants. The columns show the size of network where $n$ is the number of vertices. Numbers colored in red are worse than chance ($\frac{1}{3}$), underlined numbers show below-human performance (\textbf{Tuned} compares to human experts).}
\label{tab:merged:2}

\centering

\begin{minipage}{0.32\linewidth}
\centering
\textbf{Trained}

\smallskip

\sffamily
\tiny
\renewcommand{\arraystretch}{1.25}
\begin{tabular}{@{}ccccccccccc@{}}
\rowcolor{color1}
{\color[HTML]{F8F8F8} \textbf{$\downarrow$ Diff. $\mid$ $\rightarrow$ Size}}                                & \color[HTML]{F8F8F8}\textbf{n=10} & \color[HTML]{F8F8F8}\textbf{n=25} & \color[HTML]{F8F8F8}\textbf{n=50} & \color[HTML]{F8F8F8}\color[HTML]{F8F8F8}\textbf{Row Mean} \\
\cellcolor{color1}{\color[HTML]{F8F8F8} \textbf{0.00}} & \color{red}\underline{0.167} & \underline{0.385} & \underline{0.364} & \cellcolor{color1!30}\color{red}\underline{0.306} \\
\cellcolor{color1}{\color[HTML]{F8F8F8} \textbf{0.05}} & 0.556 & 0.714 & 0.625 & \cellcolor{color1!30}0.625 \\
\cellcolor{color1}{\color[HTML]{F8F8F8} \textbf{0.10}} & 0.571 & 0.667 & 1.0 & \cellcolor{color1!30}0.75 \\
\cellcolor{color1}{\color[HTML]{F8F8F8} \textbf{0.15}} & \underline{0.556} & 1.0 & 0.923 & \cellcolor{color1!30}0.839 \\
\cellcolor{color1}{\color[HTML]{F8F8F8} \textbf{0.20}} & \underline{0.75} & 1.0 & 1.0 & \cellcolor{color1!30}0.933 \\
\cellcolor{color1}{\color[HTML]{F8F8F8} \textbf{0.25}} & \underline{0.714} & 1.0 & 1.0 & \cellcolor{color1!30}0.905 \\
\cellcolor{color1}{\color[HTML]{F8F8F8} \textbf{0.30}} & 0.875 & 1.0 & 1.0 & \cellcolor{color1!30}0.952 \\
\cellcolor{color1}{\color[HTML]{F8F8F8} \textbf{0.35}} & 1.0 & 1.0 & 1.0 & \cellcolor{color1!30}{\phantom{0}1.0\phantom{0}} \\
\cellcolor{color1}{\color[HTML]{F8F8F8} \textbf{0.40}} & \underline{0.875} & 1.0 & 1.0 & \cellcolor{color1!30}0.958 \\
\cellcolor{color1}{\color[HTML]{F8F8F8} \textbf{Column Mean}} & \cellcolor{color1!30}\underline{0.639} & \cellcolor{color1!30}0.833 & \cellcolor{color1!30}0.847 & \cellcolor{color1!60}0.773 \\
\end{tabular}
\medskip 

\begin{tabular}{@{}ccccccccccc@{}}
\rowcolor{color2}
{\color[HTML]{F8F8F8} \textbf{$\downarrow$ Diff. $\mid$ $\rightarrow$ Size}}                                & \color[HTML]{F8F8F8}\textbf{n=10} & \color[HTML]{F8F8F8}\textbf{n=25} & \color[HTML]{F8F8F8}\textbf{n=50} & \color[HTML]{F8F8F8}\color[HTML]{F8F8F8}\textbf{Row Mean} \\
\cellcolor{color2}{\color[HTML]{F8F8F8} \textbf{0.00}} & 0.333 & \color{red}\underline{0.182} & \color{red}\underline{0.2} & \cellcolor{color2!30}\color{red}\underline{0.242} \\
\cellcolor{color2}{\color[HTML]{F8F8F8} \textbf{0.05}} & 1.0 & 0.6 & 0.667 & \cellcolor{color2!30}0.733 \\
\cellcolor{color2}{\color[HTML]{F8F8F8} \textbf{0.10}} & \underline{0.5} & 0.625 & 1.0 & \cellcolor{color2!30}0.708 \\
\cellcolor{color2}{\color[HTML]{F8F8F8} \textbf{0.15}} & 0.875 & 1.0 & 1.0 & \cellcolor{color2!30}0.958 \\
\cellcolor{color2}{\color[HTML]{F8F8F8} \textbf{0.20}} & \underline{0.75} & 1.0 & 1.0 & \cellcolor{color2!30}0.917 \\
\cellcolor{color2}{\color[HTML]{F8F8F8} \textbf{0.25}} & 1.0 & 1.0 & 1.0 & \cellcolor{color2!30}{\phantom{0}1.0\phantom{0}} \\
\cellcolor{color2}{\color[HTML]{F8F8F8} \textbf{0.30}} & 0.875 & 1.0 & 1.0 & \cellcolor{color2!30}0.958 \\
\cellcolor{color2}{\color[HTML]{F8F8F8} \textbf{0.35}} & 1.0 & 1.0 & 1.0 & \cellcolor{color2!30}{\phantom{0}1.0\phantom{0}} \\
\cellcolor{color2}{\color[HTML]{F8F8F8} \textbf{0.40}} & 1.0 & 1.0 & 1.0 & \cellcolor{color2!30}{\phantom{0}1.0\phantom{0}} \\
\cellcolor{color2}{\color[HTML]{F8F8F8} \textbf{Column Mean}} & \cellcolor{color2!30}0.778 & \cellcolor{color2!30}0.806 & \cellcolor{color2!30}0.861 & \cellcolor{color2!60}0.815 \\
\end{tabular}
\medskip 

\begin{tabular}{@{}cccccc@{}}
\rowcolor{color4}
{\color[HTML]{F8F8F8} \textbf{$\downarrow$ Diff. $\mid$ $\rightarrow$ Size}} & {\color[HTML]{F8F8F8} \textbf{n=10}} & {\color[HTML]{F8F8F8} \textbf{n=25}} & {\color[HTML]{F8F8F8} \textbf{n=50}} & {\color[HTML]{F8F8F8} \textbf{Row Mean}} \\
\cellcolor{color4}{\color[HTML]{F8F8F8}\textbf{0.00}} & 0.667 & \underline{0.538} & \underline{0.455} & \cellcolor{color4!30}0.556 \\
\cellcolor{color4}{\color[HTML]{F8F8F8}\textbf{0.05}} & 0.889 & 0.714 & 0.5   & \cellcolor{color4!30}0.708 \\
\cellcolor{color4}{\color[HTML]{F8F8F8}\textbf{0.10}} & 0.714 & 0.667 & 0.857 & \cellcolor{color4!30}0.750 \\
\cellcolor{color4}{\color[HTML]{F8F8F8}\textbf{0.15}} & \underline{0.667} & \underline{0.556} & \underline{0.692} & \cellcolor{color4!30}\underline{0.645} \\
\cellcolor{color4}{\color[HTML]{F8F8F8}\textbf{0.20}} & \underline{0.75}  & 1.0   & \underline{0.8}   & \cellcolor{color4!30}0.867 \\
\cellcolor{color4}{\color[HTML]{F8F8F8}\textbf{0.25}} & \underline{0.857} & 0.889 & 1.0   & \cellcolor{color4!30}0.905 \\
\cellcolor{color4}{\color[HTML]{F8F8F8}\textbf{0.30}} & 1.0   & 1.0   & 1.0   & \cellcolor{color4!30}1.000 \\
\cellcolor{color4}{\color[HTML]{F8F8F8}\textbf{0.35}} & \underline{0.75}  & 1.0   & 1.0   & \cellcolor{color4!30}\underline{0.917} \\
\cellcolor{color4}{\color[HTML]{F8F8F8}\textbf{0.40}} & \underline{0.875} & 1.0   & 1.0   & \cellcolor{color4!30}0.958 \\
\cellcolor{color4}{\color[HTML]{F8F8F8}\textbf{Column Mean}} & \cellcolor{color4!30}0.792 & \cellcolor{color4!30}0.792 & \cellcolor{color4!30}0.778 & \cellcolor{color4!60}0.787 \\
\end{tabular}
\medskip

\begin{tabular}{@{}ccccccccccc@{}}
\rowcolor{color3}
{\color[HTML]{F8F8F8} \textbf{$\downarrow$ Diff. $\mid$ $\rightarrow$ Size}}                                & \color[HTML]{F8F8F8}\textbf{n=10} & \color[HTML]{F8F8F8}\textbf{n=25} & \color[HTML]{F8F8F8}\textbf{n=50} & \color[HTML]{F8F8F8}\textbf{Row Mean} \\
\cellcolor{color3}{\color[HTML]{F8F8F8} \textbf{0.00}} & 0.312 & 0.552 & 0.52 & \cellcolor{color3!30}0.461 \\
\cellcolor{color3}{\color[HTML]{F8F8F8} \textbf{0.05}} & 0.472 & 0.416 & 0.472 & \cellcolor{color3!30}0.453 \\
\cellcolor{color3}{\color[HTML]{F8F8F8} \textbf{0.10}} & 0.544 & 0.648 & 0.664 & \cellcolor{color3!30}0.619 \\
\cellcolor{color3}{\color[HTML]{F8F8F8} \textbf{0.15}} & 0.704 & 0.752 & 0.784 & \cellcolor{color3!30}0.747 \\
\cellcolor{color3}{\color[HTML]{F8F8F8} \textbf{0.20}} & 0.848 & 0.8 & 0.896 & \cellcolor{color3!30}0.848 \\
\cellcolor{color3}{\color[HTML]{F8F8F8} \textbf{0.25}} & 0.864 & 0.872 & 0.904 & \cellcolor{color3!30}0.88 \\
\cellcolor{color3}{\color[HTML]{F8F8F8} \textbf{0.30}} & 0.872 & 0.912 & 0.928 & \cellcolor{color3!30}0.904 \\
\cellcolor{color3}{\color[HTML]{F8F8F8} \textbf{0.35}} & 0.936 & 0.944 & 0.92 & \cellcolor{color3!30}{0.933} \\
\cellcolor{color3}{\color[HTML]{F8F8F8} \textbf{0.40}} & 0.904 & 0.96 & 0.928 & \cellcolor{color3!30}0.931 \\
\cellcolor{color3}{\color[HTML]{F8F8F8} \textbf{Column Mean}} & \cellcolor{color3!30}0.717 & \cellcolor{color3!30}0.762 & \cellcolor{color3!30}0.78 & \cellcolor{color3!60}0.753 \\
\end{tabular}

\end{minipage}%
\begin{minipage}{0.32\linewidth}

\centering
\textbf{Untrained}

\smallskip

\sffamily
\tiny
\renewcommand{\arraystretch}{1.25}
\begin{tabular}{@{}ccccccccccc@{}}
\rowcolor{color1}
{\color[HTML]{F8F8F8} \textbf{$\downarrow$ Diff. $\mid$ $\rightarrow$ Size}}                                & \color[HTML]{F8F8F8}\textbf{n=10} & \color[HTML]{F8F8F8}\textbf{n=25} & \color[HTML]{F8F8F8}\textbf{n=50} & \color[HTML]{F8F8F8}\textbf{Row Mean} \\
\cellcolor{color1}{\color[HTML]{F8F8F8} \textbf{0.00}} & \color{red}\underline{0.25} & \color{red}\underline{0.154} & \color{red}\underline{0.273} & \cellcolor{color1!30}\color{red}\underline{0.222} \\
\cellcolor{color1}{\color[HTML]{F8F8F8} \textbf{0.05}} & 0.556 & 0.714 & 0.875 & \cellcolor{color1!30}0.708 \\
\cellcolor{color1}{\color[HTML]{F8F8F8} \textbf{0.10}} & \underline{0.429} & 0.833 & 1.0 & \cellcolor{color1!30}0.75 \\
\cellcolor{color1}{\color[HTML]{F8F8F8} \textbf{0.15}} & 0.667 & 0.778 & 0.846 & \cellcolor{color1!30}0.774 \\
\cellcolor{color1}{\color[HTML]{F8F8F8} \textbf{0.20}} & 0.75 & 1.0 & 1.0 & \cellcolor{color1!30}0.933 \\
\cellcolor{color1}{\color[HTML]{F8F8F8} \textbf{0.25}} & \underline{0.714} & 1.0 & 1.0 & \cellcolor{color1!30}0.905 \\
\cellcolor{color1}{\color[HTML]{F8F8F8} \textbf{0.30}} & 0.75 & 1.0 & 1.0 & \cellcolor{color1!30}0.905 \\
\cellcolor{color1}{\color[HTML]{F8F8F8} \textbf{0.35}} & \underline{0.75} & 1.0 & 1.0 & \cellcolor{color1!30}0.917 \\
\cellcolor{color1}{\color[HTML]{F8F8F8} \textbf{0.40}} & 0.875 & 1.0 & 1.0 & \cellcolor{color1!30}0.958 \\
\cellcolor{color1}{\color[HTML]{F8F8F8} \textbf{Column Mean}} & \cellcolor{color1!30}\underline{0.611} & \cellcolor{color1!30}0.778 & \cellcolor{color1!30}0.847 & \cellcolor{color1!60}0.745 \\
\end{tabular}
\medskip 

\begin{tabular}{@{}ccccccccccc@{}}
\rowcolor{color2}
{\color[HTML]{F8F8F8} \textbf{$\downarrow$ Diff. $\mid$ $\rightarrow$ Size}}                                & \color[HTML]{F8F8F8}\textbf{n=10} & \color[HTML]{F8F8F8}\textbf{n=25} & \color[HTML]{F8F8F8}\textbf{n=50} & \color[HTML]{F8F8F8}\textbf{Row Mean} \\
\cellcolor{color2}{\color[HTML]{F8F8F8} \textbf{0.00}} & \underline{{0.167}} & \color{red}\underline{0.091} & \color{red}\underline{0.2} & \cellcolor{color2!30}\color{red}\underline{0.152} \\
\cellcolor{color2}{\color[HTML]{F8F8F8} \textbf{0.05}} & 0.75 & 0.6 & 1.0 & \cellcolor{color2!30}0.8 \\
\cellcolor{color2}{\color[HTML]{F8F8F8} \textbf{0.10}} & 0.5 & 0.875 & 0.875 & \cellcolor{color2!30}0.75 \\
\cellcolor{color2}{\color[HTML]{F8F8F8} \textbf{0.15}} & 1.0 & 1.0 & 1.0 & \cellcolor{color2!30}1.0 \\
\cellcolor{color2}{\color[HTML]{F8F8F8} \textbf{0.20}} & 0.875 & 1.0 & 0.75 & \cellcolor{color2!30}0.875 \\
\cellcolor{color2}{\color[HTML]{F8F8F8} \textbf{0.25}} & 1.0 & 1.0 & 1.0 & \cellcolor{color2!30}1.0 \\
\cellcolor{color2}{\color[HTML]{F8F8F8} \textbf{0.30}} & 1.0 & 1.0 & 1.0 & \cellcolor{color2!30}1.0 \\
\cellcolor{color2}{\color[HTML]{F8F8F8} \textbf{0.35}} & 1.0 & 1.0 & 1.0 & \cellcolor{color2!30}1.0 \\
\cellcolor{color2}{\color[HTML]{F8F8F8} \textbf{0.40}} & 1.0 & 1.0 & 1.0 & \cellcolor{color2!30}1.0 \\
\cellcolor{color2}{\color[HTML]{F8F8F8} \textbf{Column Mean}} & \cellcolor{color2!30}0.778 & \cellcolor{color2!30}0.819 & \cellcolor{color2!30}0.847 & \cellcolor{color2!60}0.815 \\
\end{tabular}
\medskip 

\begin{tabular}{@{}cccccc@{}}
\rowcolor{color4n}
{\color[HTML]{F8F8F8} \textbf{$\downarrow$ Diff. $\mid$ $\rightarrow$ Size}} & {\color[HTML]{F8F8F8} \textbf{n=10}} & {\color[HTML]{F8F8F8} \textbf{n=25}} & {\color[HTML]{F8F8F8} \textbf{n=50}} & {\color[HTML]{F8F8F8} \textbf{Row Mean}} \\
\cellcolor{color4n}{\color[HTML]{F8F8F8}\textbf{0.00}} & \color{red}\underline{0.25}  & \color{red}\underline{0.308} & \underline{0.455} & \cellcolor{color4n!30}0.333 \\
\cellcolor{color4n}{\color[HTML]{F8F8F8}\textbf{0.05}} & 0.889 & 0.857 & 0.5   & \cellcolor{color4n!30}0.750 \\
\cellcolor{color4n}{\color[HTML]{F8F8F8}\textbf{0.10}} & 1.0   & 0.833 & 0.714 & \cellcolor{color4n!30}0.850 \\
\cellcolor{color4n}{\color[HTML]{F8F8F8}\textbf{0.15}} & \underline{0.444} & 0.667 & 0.846 & \cellcolor{color4n!30}0.677 \\
\cellcolor{color4n}{\color[HTML]{F8F8F8}\textbf{0.20}} & 1.0   & 1.0 & 1.0   & \cellcolor{color4n!30}1.000 \\
\cellcolor{color4n}{\color[HTML]{F8F8F8}\textbf{0.25}} & 0.857 & 0.889 & 1.0   & \cellcolor{color4n!30}0.905 \\
\cellcolor{color4n}{\color[HTML]{F8F8F8}\textbf{0.30}} & 1.0   & 1.0   & 1.0   & \cellcolor{color4n!30}1.000 \\
\cellcolor{color4n}{\color[HTML]{F8F8F8}\textbf{0.35}} & 0.875 & 1.0   & 1.0   & \cellcolor{color4n!30}0.958 \\
\cellcolor{color4n}{\color[HTML]{F8F8F8}\textbf{0.40}} & 0.875 & 1.0   & 1.0   & \cellcolor{color4n!30}0.958 \\
\cellcolor{color4n}{\color[HTML]{F8F8F8}\textbf{Column Mean}} & \cellcolor{color4n!30}0.750 & \cellcolor{color4n!30}0.792 & \cellcolor{color4n!30}0.806 & \cellcolor{color4n!60}0.782 \\
\end{tabular}

\medskip

\begin{tabular}{@{}ccccccccccc@{}}
\rowcolor{color3}
{\color[HTML]{F8F8F8} \textbf{$\downarrow$ Diff. $\mid$ $\rightarrow$ Size}}                                & \color[HTML]{F8F8F8}\textbf{n=10} & \color[HTML]{F8F8F8}\textbf{n=25} & \color[HTML]{F8F8F8}\textbf{n=50} & \color[HTML]{F8F8F8}\textbf{Row Mean} \\
\cellcolor{color3}{\color[HTML]{F8F8F8} \textbf{0.00}} & 0.464 & 0.712 & 0.68 & \cellcolor{color3!30}0.619 \\
\cellcolor{color3}{\color[HTML]{F8F8F8} \textbf{0.05}} & 0.36 & 0.304 & 0.288 & \cellcolor{color3!30}0.317 \\
\cellcolor{color3}{\color[HTML]{F8F8F8} \textbf{0.10}} & 0.464 & 0.592 & 0.48 & \cellcolor{color3!30}0.512 \\
\cellcolor{color3}{\color[HTML]{F8F8F8} \textbf{0.15}} & 0.624 & 0.656 & 0.552 & \cellcolor{color3!30}0.611 \\
\cellcolor{color3}{\color[HTML]{F8F8F8} \textbf{0.20}} & 0.68 & 0.816 & 0.696 & \cellcolor{color3!30}0.731 \\
\cellcolor{color3}{\color[HTML]{F8F8F8} \textbf{0.25}} & 0.736 & 0.856 & 0.664 & \cellcolor{color3!30}0.752 \\
\cellcolor{color3}{\color[HTML]{F8F8F8} \textbf{0.30}} & 0.736 & 0.848 & 0.76 & \cellcolor{color3!30}0.781 \\
\cellcolor{color3}{\color[HTML]{F8F8F8} \textbf{0.35}} & 0.88 & 0.912 & 0.768 & \cellcolor{color3!30}0.853 \\
\cellcolor{color3}{\color[HTML]{F8F8F8} \textbf{0.40}} & 0.84 & 0.888 & 0.824 & \cellcolor{color3!30}0.851 \\
\cellcolor{color3}{\color[HTML]{F8F8F8} \textbf{Column Mean}} & \cellcolor{color3!30}0.643 & \cellcolor{color3!30}0.732 & \cellcolor{color3!30}0.635 & \cellcolor{color3!60}0.67 \\
\end{tabular}

\end{minipage}%
\begin{minipage}{0.32\linewidth}

\centering
\textbf{Expert}

\smallskip

\sffamily
\footnotesize
\renewcommand{\arraystretch}{1.25}
\tiny
\begin{tabular}{@{}ccccccccccc@{}}
\rowcolor{color1}
{\color[HTML]{F8F8F8} \textbf{$\downarrow$ Diff. $\mid$ $\rightarrow$ Size}}                                & \color[HTML]{F8F8F8}\textbf{n=10} & \color[HTML]{F8F8F8}\textbf{n=25} & \color[HTML]{F8F8F8}\textbf{n=50} & \color[HTML]{F8F8F8}\textbf{Row Mean} \\
\cellcolor{color1}{\color[HTML]{F8F8F8} \textbf{0.00}} & \color{red}\underline{0.167} & \color{red}\underline{0.308} & \underline{0.455} & \cellcolor{color1!30}\color{red}\underline{0.306} \\
\cellcolor{color1}{\color[HTML]{F8F8F8} \textbf{0.05}} & \underline{0.667} & 0.571 & 0.5 & \cellcolor{color1!30}0.583 \\
\cellcolor{color1}{\color[HTML]{F8F8F8} \textbf{0.10}} & \underline{0.429} & \underline{0.5} & \underline{0.429} & \cellcolor{color1!30}\underline{0.45} \\
\cellcolor{color1}{\color[HTML]{F8F8F8} \textbf{0.15}} & 0.778 & \underline{0.667} & \underline{0.692} & \cellcolor{color1!30}\underline{0.71} \\
\cellcolor{color1}{\color[HTML]{F8F8F8} \textbf{0.20}} & 1.0 & 0.833 & 0.8 & \cellcolor{color1!30}0.867 \\
\cellcolor{color1}{\color[HTML]{F8F8F8} \textbf{0.25}} & \underline{0.857} & 0.889 & \underline{0.8} & \cellcolor{color1!30}\underline{0.857} \\
\cellcolor{color1}{\color[HTML]{F8F8F8} \textbf{0.30}} & \underline{0.75} & 1.0 & 1.0 & \cellcolor{color1!30}0.905 \\
\cellcolor{color1}{\color[HTML]{F8F8F8} \textbf{0.35}} & \underline{0.625} & 1.0 & 1.0 & \cellcolor{color1!30}\underline{0.875} \\
\cellcolor{color1}{\color[HTML]{F8F8F8} \textbf{0.40}} & 1.0 & 1.0 & 1.0 & \cellcolor{color1!30}{\phantom{0}1.0\phantom{0}} \\
\cellcolor{color1}{\color[HTML]{F8F8F8} \textbf{Column Mean}} & \cellcolor{color1!30}\underline{0.653} & \cellcolor{color1!30}\underline{0.722} & \cellcolor{color1!30}\underline{0.722} & \cellcolor{color1!60}\underline{0.699} \\
\end{tabular}
\medskip 

\begin{tabular}{@{}ccccccccccc@{}}
\rowcolor{color2}
{\color[HTML]{F8F8F8} \textbf{$\downarrow$ Diff. $\mid$ $\rightarrow$ Size}}                                & \color[HTML]{F8F8F8}\textbf{n=10} & \color[HTML]{F8F8F8}\textbf{n=25} & \color[HTML]{F8F8F8}\textbf{n=50} & \color[HTML]{F8F8F8}\textbf{Row Mean} \\
\cellcolor{color2}{\color[HTML]{F8F8F8} \textbf{0.00}} & \underline{0.417} & 0.545 & \underline{0.6} & \cellcolor{color2!30}\underline{0.515} \\
\cellcolor{color2}{\color[HTML]{F8F8F8} \textbf{0.05}} & 0.75 & \underline{0.4} & \underline{0.333} & \cellcolor{color2!30}\underline{0.467} \\
\cellcolor{color2}{\color[HTML]{F8F8F8} \textbf{0.10}} & \underline{0.375} & \underline{0.375} & \underline{0.375} & \cellcolor{color2!30}\underline{0.375} \\
\cellcolor{color2}{\color[HTML]{F8F8F8} \textbf{0.15}} & \underline{0.5} & \underline{0.75} & \underline{0.375} & \cellcolor{color2!30}\underline{0.542} \\
\cellcolor{color2}{\color[HTML]{F8F8F8} \textbf{0.20}} & \underline{0.5} & \underline{0.75} & \underline{0.625} & \cellcolor{color2!30}\underline{0.625} \\
\cellcolor{color2}{\color[HTML]{F8F8F8} \textbf{0.25}} & \underline{0.875} & \underline{0.875} & \underline{0.75} & \cellcolor{color2!30}\underline{0.833} \\
\cellcolor{color2}{\color[HTML]{F8F8F8} \textbf{0.30}} & \underline{0.75} & 1.0 & \underline{0.875} & \cellcolor{color2!30}\underline{0.875} \\
\cellcolor{color2}{\color[HTML]{F8F8F8} \textbf{0.35}} & \underline{0.75} & 1.0 & \underline{1.0} & \cellcolor{color2!30}\underline{0.917} \\
\cellcolor{color2}{\color[HTML]{F8F8F8} \textbf{0.40}} & 1.0 & 1.0 & 0.75 & \cellcolor{color2!30}\underline{0.917} \\
\cellcolor{color2}{\color[HTML]{F8F8F8} \textbf{Column Mean}} & \cellcolor{color2!30}\underline{0.639} & \cellcolor{color2!30}\underline{0.75} & \cellcolor{color2!30}\underline{0.639} & \cellcolor{color2!60}\underline{0.676} \\
\end{tabular}
\medskip 

\begin{tabular}{@{}cccccc@{}}
\rowcolor{color4}
{\color[HTML]{F8F8F8} \textbf{$\downarrow$ Diff. $\mid$ $\rightarrow$ Size}} & {\color[HTML]{F8F8F8} \textbf{n=10}} & {\color[HTML]{F8F8F8} \textbf{n=25}} & {\color[HTML]{F8F8F8} \textbf{n=50}} & {\color[HTML]{F8F8F8} \textbf{Row Mean}} \\
\cellcolor{color4}{\color[HTML]{F8F8F8}\textbf{0.00}} & 0.75  & 0.615 & 0.818 & \cellcolor{color4!30}0.722 \\
\cellcolor{color4}{\color[HTML]{F8F8F8}\textbf{0.05}} & \color{red}\underline{0.222} & \color{red}\underline{0.143} & \underline{0.375} & \cellcolor{color4!30}\color{red}\underline{0.250} \\
\cellcolor{color4}{\color[HTML]{F8F8F8}\textbf{0.10}} & \color{red}\underline{0.286} & \color{red}\underline{0.167} & \color{red}\underline{0.286} & \cellcolor{color4!30}\color{red}\underline{0.250} \\
\cellcolor{color4}{\color[HTML]{F8F8F8}\textbf{0.15}} & \color{red}\underline{0.111} & \color{red}\underline{0.333} & \color{red}\underline{0.231} & \cellcolor{color4!30}\color{red}\underline{0.226} \\
\cellcolor{color4}{\color[HTML]{F8F8F8}\textbf{0.20}} & \color{red}\underline{0.25}  & \underline{0.833} & \underline{0.4}   & \cellcolor{color4!30}\underline{0.533} \\
\cellcolor{color4}{\color[HTML]{F8F8F8}\textbf{0.25}} & \underline{0.571} & \underline{0.778} & \underline{0.4}   & \cellcolor{color4!30}\underline{0.619} \\
\cellcolor{color4}{\color[HTML]{F8F8F8}\textbf{0.30}} & \underline{0.375} & 1.0   & 1.0   & \cellcolor{color4!30}\underline{0.762} \\
\cellcolor{color4}{\color[HTML]{F8F8F8}\textbf{0.35}} & \underline{0.375} & 1.0   & \underline{0.625} & \cellcolor{color4!30}\underline{0.667} \\
\cellcolor{color4}{\color[HTML]{F8F8F8}\textbf{0.40}} & \underline{0.75}  & \underline{0.875} & \underline{0.875} & \cellcolor{color4!30}\underline{0.833} \\
\cellcolor{color4}{\color[HTML]{F8F8F8}\textbf{Column Mean}} & \cellcolor{color4!30}\underline{0.431} & \cellcolor{color4!30}\underline{0.639} & \cellcolor{color4!30}\underline{0.556} & \cellcolor{color4!60}\underline{0.542} \\
\end{tabular}
\medskip

\begin{tabular}{@{}ccccccccccc@{}}
\rowcolor{color3}
{\color[HTML]{F8F8F8} \textbf{$\downarrow$ Diff. $\mid$ $\rightarrow$ Size}}                                & \color[HTML]{F8F8F8}\textbf{n=10} & \color[HTML]{F8F8F8}\textbf{n=25} & \color[HTML]{F8F8F8}\textbf{n=50} & \color[HTML]{F8F8F8}\textbf{Row Mean} \\
\cellcolor{color3}{\color[HTML]{F8F8F8} \textbf{0.00}} & 0.444 & 0.489 & 0.711 & \cellcolor{color3!30}0.548 \\
\cellcolor{color3}{\color[HTML]{F8F8F8} \textbf{0.05}} & 0.444 & 0.489 & 0.489 & \cellcolor{color3!30}0.474 \\
\cellcolor{color3}{\color[HTML]{F8F8F8} \textbf{0.10}} & 0.444 & 0.778 & 0.644 & \cellcolor{color3!30}0.622 \\
\cellcolor{color3}{\color[HTML]{F8F8F8} \textbf{0.15}} & 0.778 & 0.711 & 0.711 & \cellcolor{color3!30}0.733 \\
\cellcolor{color3}{\color[HTML]{F8F8F8} \textbf{0.20}} & 0.778 & 0.867 & 0.822 & \cellcolor{color3!30}0.822 \\
\cellcolor{color3}{\color[HTML]{F8F8F8} \textbf{0.25}} & 0.911 & 0.889 & 0.956 & \cellcolor{color3!30}0.919 \\
\cellcolor{color3}{\color[HTML]{F8F8F8} \textbf{0.30}} & 0.822 & 0.889 & 0.978 & \cellcolor{color3!30}0.896 \\
\cellcolor{color3}{\color[HTML]{F8F8F8} \textbf{0.35}} & 0.911 & 0.956 & 0.956 & \cellcolor{color3!30}0.941 \\
\cellcolor{color3}{\color[HTML]{F8F8F8} \textbf{0.40}} & 0.978 & 1.0 & 1.0 & \cellcolor{color3!30}0.993 \\
\cellcolor{color3}{\color[HTML]{F8F8F8} \textbf{Column Mean}} & \cellcolor{color3!30}0.723 & \cellcolor{color3!30}0.785 & \cellcolor{color3!30}0.807 & \cellcolor{color3!60}0.772 \\
\end{tabular}
    
\end{minipage}

\vspace{1.5em}
\centering
\textbf{Tuned}

\smallskip

\begin{minipage}{0.32\linewidth}
\centering
\sffamily
\footnotesize
\renewcommand{\arraystretch}{1.25}
\tiny
\begin{tabular}{@{}ccccccccccc@{}}
\rowcolor{color1}
{\color[HTML]{F8F8F8} \textbf{$\downarrow$ Diff. $\mid$ $\rightarrow$ Size}}                                & \textbf{\color[HTML]{F8F8F8}n=10} & \textbf{\color[HTML]{F8F8F8}n=25} & \textbf{\color[HTML]{F8F8F8}n=50} & \textbf{\color[HTML]{F8F8F8}Row Mean} \\
\cellcolor{color1}{\color[HTML]{F8F8F8} \textbf{0.00}} & \color{red}\underline{0.167} & \color{red}\underline{0.182} & \color{red}\underline{0.2} & \cellcolor{color1!30}\color{red}\underline{0.182} \\
\cellcolor{color1}{\color[HTML]{F8F8F8} \textbf{0.05}} & 0.75 & 0.8 & 0.5 & \cellcolor{color1!30}0.667 \\
\cellcolor{color1}{\color[HTML]{F8F8F8} \textbf{0.10}} & \underline{0.375} & 0.875 & 1.0 & \cellcolor{color1!30}0.75 \\
\cellcolor{color1}{\color[HTML]{F8F8F8} \textbf{0.15}} & \underline{0.75} & 1.0 & 1.0 & \cellcolor{color1!30}0.917 \\
\cellcolor{color1}{\color[HTML]{F8F8F8} \textbf{0.20}} & \underline{0.625} & 1.0 & 1.0 & \cellcolor{color1!30}0.875 \\
\cellcolor{color1}{\color[HTML]{F8F8F8} \textbf{0.25}} & \underline{0.875} & 1.0 & 1.0 & \cellcolor{color1!30}0.958 \\
\cellcolor{color1}{\color[HTML]{F8F8F8} \textbf{0.30}} & 0.875 & 1.0 & 1.0 & \cellcolor{color1!30}0.958 \\
\cellcolor{color1}{\color[HTML]{F8F8F8} \textbf{0.35}} & \underline{0.875} & 1.0 & 1.0 & \cellcolor{color1!30}0.958 \\
\cellcolor{color1}{\color[HTML]{F8F8F8} \textbf{0.40}} & 1.0 & 1.0 & 1.0 & \cellcolor{color1!30}{\phantom{0}1.0\phantom{0}} \\
\cellcolor{color1}{\color[HTML]{F8F8F8} \textbf{Column Mean}} & \cellcolor{color1!30}\underline{0.667} & \cellcolor{color1!30}0.847 & \cellcolor{color1!30}0.847 & \cellcolor{color1!60}0.787 \\
\end{tabular}
\end{minipage}%
\begin{minipage}{0.32\linewidth}
\centering
\sffamily
\footnotesize
\renewcommand{\arraystretch}{1.25}
\tiny
\begin{tabular}{@{}ccccccccccc@{}}
\rowcolor{color2}
{\color[HTML]{F8F8F8} \textbf{$\downarrow$ Diff. $\mid$ $\rightarrow$ Size}}                                & \textbf{\color[HTML]{F8F8F8}n=10} & \textbf{\color[HTML]{F8F8F8}n=25} & \textbf{\color[HTML]{F8F8F8}n=50} & \textbf{\color[HTML]{F8F8F8}Row Mean} \\
\cellcolor{color2}{\color[HTML]{F8F8F8} \textbf{0.00}} & \underline{0.333} & \underline{0.364} & \underline{0.4} & \cellcolor{color2!30}\underline{0.364} \\
\cellcolor{color2}{\color[HTML]{F8F8F8} \textbf{0.05}} & 1.0 & 0.6 & 0.5 & \cellcolor{color2!30}0.667 \\
\cellcolor{color2}{\color[HTML]{F8F8F8} \textbf{0.10}} & 0.5 & 0.875 & 0.625 & \cellcolor{color2!30}0.667 \\
\cellcolor{color2}{\color[HTML]{F8F8F8} \textbf{0.15}} & \underline{0.75} & 1.0 & 0.875 & \cellcolor{color2!30}0.875 \\
\cellcolor{color2}{\color[HTML]{F8F8F8} \textbf{0.20}} & 1.0 & 1.0 & 1.0 & \cellcolor{color2!30}{\phantom{0}1.0\phantom{0}} \\
\cellcolor{color2}{\color[HTML]{F8F8F8} \textbf{0.25}} & 1.0 & 1.0 & 1.0 & \cellcolor{color2!30}{\phantom{0}1.0\phantom{0}} \\
\cellcolor{color2}{\color[HTML]{F8F8F8} \textbf{0.30}} & 1.0 & 1.0 & 1.0 & \cellcolor{color2!30}{\phantom{0}1.0\phantom{0}} \\
\cellcolor{color2}{\color[HTML]{F8F8F8} \textbf{0.35}} & 1.0 & 1.0 & 1.0 & \cellcolor{color2!30}{\phantom{0}1.0\phantom{0}} \\
\cellcolor{color2}{\color[HTML]{F8F8F8} \textbf{0.40}} & 1.0 & 1.0 & 1.0 & \cellcolor{color2!30}{\phantom{0}1.0\phantom{0}} \\
\cellcolor{color2}{\color[HTML]{F8F8F8} \textbf{Column Mean}} & \cellcolor{color2!30}\underline{0.806} & \cellcolor{color2!30}0.861 & \cellcolor{color2!30}0.819 & \cellcolor{color2!60}0.829 \\
\end{tabular}
\end{minipage}%
\begin{minipage}{0.32\linewidth}
\centering
\sffamily
\footnotesize
\renewcommand{\arraystretch}{1.25}
\tiny
\begin{tabular}{@{}cccccc@{}}
\rowcolor{color4}
{\color[HTML]{F8F8F8} \textbf{$\downarrow$ Diff. $\mid$ $\rightarrow$ Size}} & {\color[HTML]{F8F8F8} \textbf{n=10}} & {\color[HTML]{F8F8F8} \textbf{n=25}} & {\color[HTML]{F8F8F8} \textbf{n=50}} & {\color[HTML]{F8F8F8} \textbf{Row Mean}} \\
\cellcolor{color4}{\color[HTML]{F8F8F8}\textbf{0.00}} & 0.667 & 0.615 & \underline{0.545} & \cellcolor{color4!30}0.611 \\
\cellcolor{color4}{\color[HTML]{F8F8F8}\textbf{0.05}} & 0.556 & 0.714 & 0.5   & \cellcolor{color4!30}0.583 \\
\cellcolor{color4}{\color[HTML]{F8F8F8}\textbf{0.10}} & \textcolor{red}{\underline{0.286}} & \underline{0.5}   & 0.714 & \cellcolor{color4!30}\underline{0.500} \\
\cellcolor{color4}{\color[HTML]{F8F8F8}\textbf{0.15}} & \underline{0.667} & 0.889 & 0.769 & \cellcolor{color4!30}0.774 \\
\cellcolor{color4}{\color[HTML]{F8F8F8}\textbf{0.20}} & 1.0   & 1.0   & \underline{0.8}   & \cellcolor{color4!30}0.933 \\
\cellcolor{color4}{\color[HTML]{F8F8F8}\textbf{0.25}} & 1.0   & 0.889 & 1.0   & \cellcolor{color4!30}0.952 \\
\cellcolor{color4}{\color[HTML]{F8F8F8}\textbf{0.30}} & 1.0   & 1.0   & 1.0   & \cellcolor{color4!30}1.000 \\
\cellcolor{color4}{\color[HTML]{F8F8F8}\textbf{0.35}} & \underline{0.625} & 1.0   & 1.0   & \cellcolor{color4!30}\underline{0.875} \\
\cellcolor{color4}{\color[HTML]{F8F8F8}\textbf{0.40}} & 1.0   & 1.0   & 1.0   & \cellcolor{color4!30}1.000 \\
\cellcolor{color4}{\color[HTML]{F8F8F8}\textbf{Column Mean}} & \cellcolor{color4!30}0.736 & \cellcolor{color4!30}0.833 & \cellcolor{color4!30}\underline{0.792} & \cellcolor{color4!60}0.787 \\
\end{tabular}
\end{minipage}

\end{table*}

\end{document}